\documentclass[12pt]{elsarticle}  
\usepackage[utf8]{inputenc}
\usepackage{latexsym}
\usepackage{amssymb}
\usepackage{amsmath}
\usepackage{graphicx}
\usepackage{subfigure}
\usepackage{color}
\usepackage{hyperref}

\begin{document}
	
	\begin{frontmatter}
		
		
		
		\title{Unitarity and Finkelstein-Kajantie problem \\in diffraction hadron production}
		
		
		\author{E. Martynov}
		\ead{martynov@bitp.kiev.ua}
		\author{G. Tersimonov}
		\ead{tersimonov@bitp.kiev.ua} 
		\address{Bogolyubov Institute for Theoretical Physics of NAS of Ukraine, Metrologichna Str., 14~b, Kiev-143}

	\begin{abstract}
		The diffraction production of many hadron showers separated by large rapidity gaps, when calculated within the standard pomeron approach, lead to cross sections rising much faster than Froissart-Martin bound. This is the point of Finkelstein-Kajantie problem. We consider the unitarization procedure based on  Dyson-Schwinger equations with input froissaron  propagators and 3-froissaron vertex (3f-vertex) depending on angular momenta of froissarons in it. The developed diffraction production model allows to resolve Finkelstein-Kajantie problem.

	\end{abstract}

		\begin{keyword}
	Pomeron, Froissaron, Dyson-Schwinger Equations, Finkelstein-Kajantie Problem
		\end{keyword}

\end{frontmatter}

	\section{Introduction}\label{sec:Introd}
	A problem of the unitarity violation in the pomeron models is well known since 1960s \cite{Verdiev, Finkelstein,G-M}. It is more related to a 3-pomeron interaction vertex rather than to a possible large intercept ($\alpha(0)>1$) of a bare pomeron. If the pomeron trajectory is linear, so $\alpha(t)=\alpha(0)+\alpha't$ with $\alpha(0)=1$, and the 3P-vertex, $r_{3P}$, is constant, then total hadron cross section does not depend asymptotically on hadron energy, $\sigma_{t}(s)\propto (s/s_{0})^{\alpha_(0)-1}={\rm const}$ ($s_{0}\sim 1 {\rm GeV}^{2}$).  At the same time the contribution of diffraction production of  high effective mass hadron showers, separated by the large enough rapidity gaps,  to $\sigma_{t}(s)$  rise with energy vigorously demonstrating inconsistency of this simplified scheme (\cite{Gribov}, \cite{Kaid-T-M},  \cite{Kaidalov} and references therein).
	\begin{figure}[h]
		\begin{center}
			\includegraphics[scale=0.5]{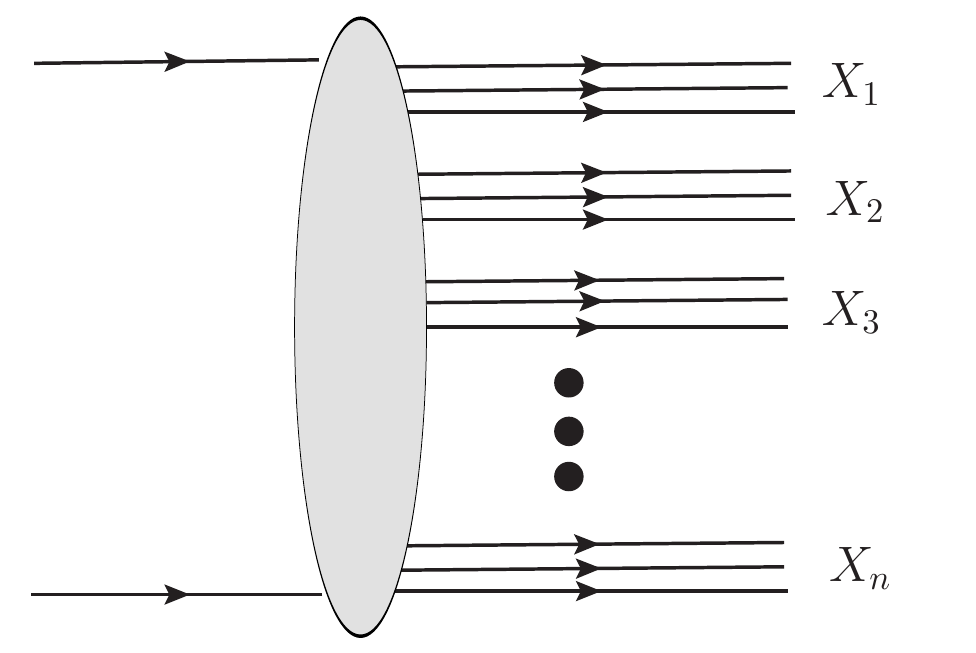}
			\caption{Diffraction production of $n$ hadron showers}
			\label{fig:n showers}
		\end{center}
	\end{figure}
	
\begin{figure}[h]
		\begin{center}
			\includegraphics[scale=0.5]{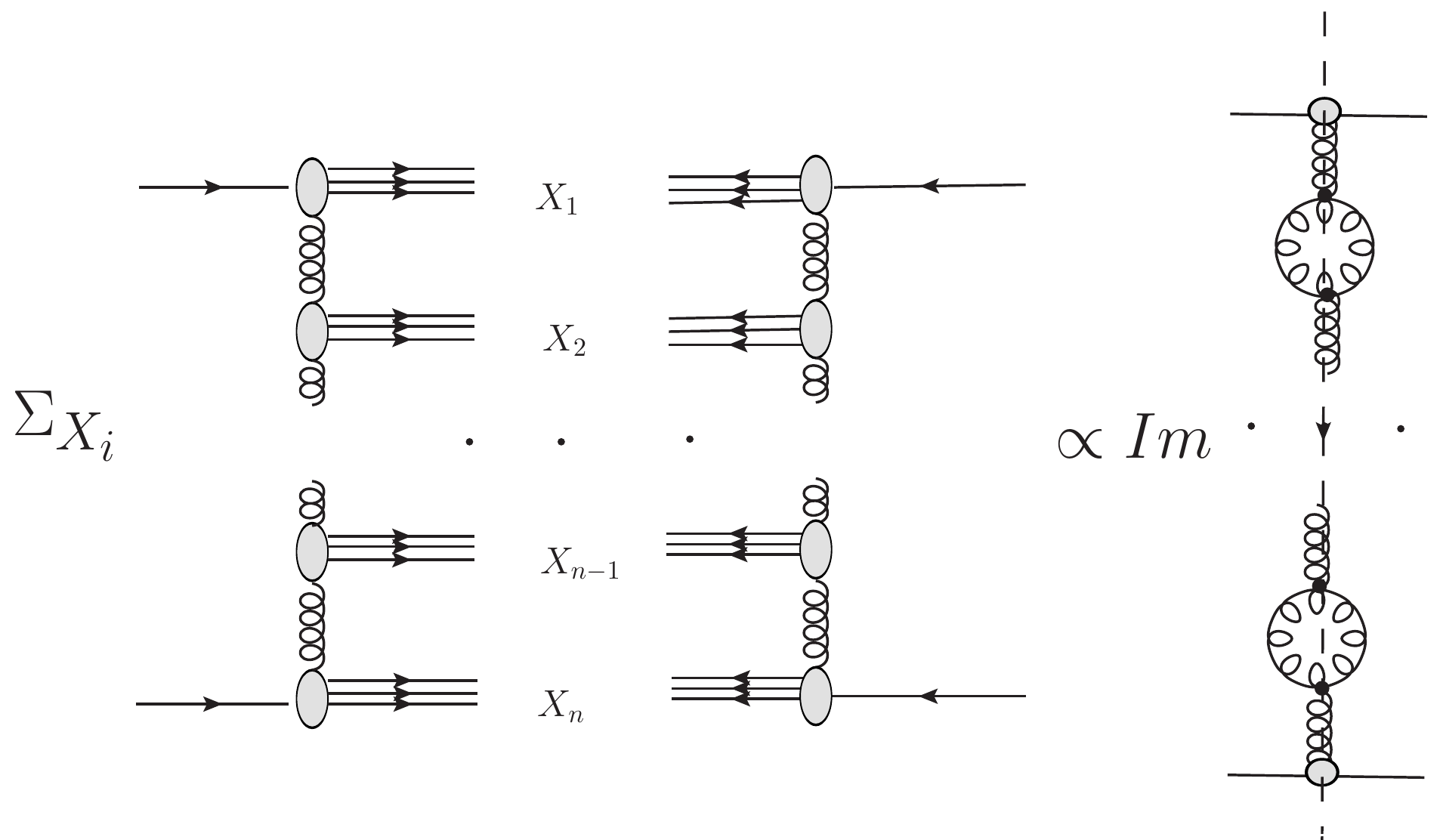}
			\caption{Cross section of a diffraction production of $n$ hadron showers as (n-1)-loop diagram}
			\label{fig:n loops}
		\end{center}
	\end{figure}

	This process is pictured by the loop diagrams shown on the Fig. \ref {fig:n loops}. The equivalence of left and right diagrams is a consequence of the generalized optical theorem.
	
	Total cross section of the $n$ showers production is determined by imaginary part of the loop diagram which has in  $j=\omega+1, t$-representation the following form 
	\begin{equation}\label{eq:n-loop}
		\phi (\omega ,0)\propto (r_{3P}^{2})^{n-1}\frac{(\ln\omega/\omega)^{n-1}}{\omega}.
	\end{equation}
	One can obtain from this equation that at $s\to \infty$ 
\begin{equation}\label{eq:n-loop-s}
	\sigma^{n}_{diifr}(s))\propto (r_{3P}^{2}\ln(s/s_{0})\ln\ln(s/s_{0}))^{n-1}, \quad s_0\sim 1\,\, {\rm GeV}^2
\end{equation} 
which comes in severe contradiction with the Froissart-Martin bound \cite{Froissart, Martin,  Mar-Luk}. This is the essence of Finkelstein-Kajantie problem \cite{Finkelstein}, \cite{ABSW}, \cite{BW}.
	
	In 1970s, Cardy proposed to consider additional pomeron re-scatterings which had to screen large rapidity gaps \cite{Cardy}. It was believed that in a black disc limit such a screening can put the cross section back to Froissart-Martin unitarity bound \cite{GLM}, \cite{KMR-1}. Unfortunately, any eikonal type  screening appears to be not enough \cite{MS}, \cite{Mart-Ters}, at least for the simplest input contribution to the amplitudes of  SDD (Single Diffraction Dissociation), CDP (central Diffraction Production) or DDD (double Diffraction Dissociation). In the paper \cite{KMR-3}  the differential cross sections have been written in a general form taking into account enhanced reggeon diagrams. However, it is most likely, that the explicit final result for integrated (over rapidity and impact parameter vari\-ab\-les) cross section cannot be obtained in analytical form. Another way to fix the problem, namely assumption that 3P-vertex  depend on $t$, $r_{3P}(t)\propto t$ at $t\to 0$ \cite{Kaidalov,BTW} does not supported by the data at high energies. 

	The pomeron with $\alpha(t)=1+\varepsilon+\alpha't$ as input in the eikonal (\cite{Collins} and references to earlier papers therein, \cite{Cheng}), quasieikonal \cite{T-M},  $U$-matrix \cite{T-T} unitarization and their generalization \cite{CSP}  lead to the elastic scattering amplitude which does not violate the Froissart-Martin limit for total cross section. Namely, unitarization lead to $\sigma_{t}\approx 8\pi \varepsilon \alpha' \ln^{2}(s/s_{0})$. Such an amplitude (in a simplified form at $s\to\infty$) can be  represented in impact parameter representation  $H(s,b)=g\Theta(R(s)-b)$ where $R(s)\propto \ln(s/s_{0})$  and $g\leq 1$. In the $\omega$-representation this amplitude is not a pole, it looks like a pair of two complex branch points colliding at $t=0$.
	
	\begin{equation}\label{eq:tripole}
		\phi(\omega,t)\propto (\omega^{2}+a^{2}\vec{q}\, ^{2})^{-3/2}, \qquad \omega=j-1, \quad \vec{q}\, ^{2}=-t
	\end{equation}
	
	\begin{figure}[!h]
		\centering
		\includegraphics[width=0.6\linewidth]{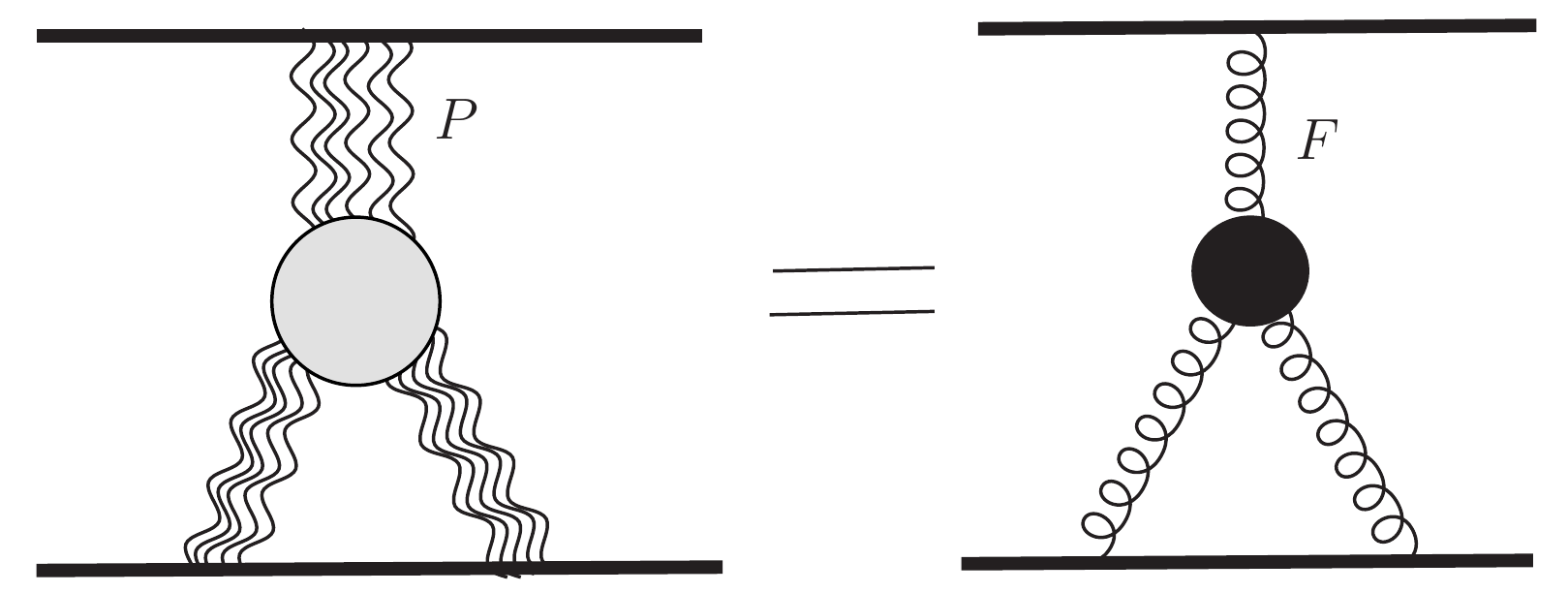}
		\caption{3f-vertex diagram as a sum of the single pomeron pole re-scatterings}
		\label{fig:3p-3f}
	\end{figure}
	
	Such a pomeron was called \textit{froissaron} \cite{DKLT-M, DT-M} because it saturates the Froissart-Martin bound (in a functional form), i.e. provides the maximality of strong interactions.
	
	By introducing the proper 3-froissaron vertex (Fig. \ref{fig:3p-3f}), which can/must depend on spatial and angular momenta of  froissarons ln it, one can hope to extend an unitarization to the shower production.
	
	Anyway, there are two possibilities:  either we start from the single pomeron pole with the intercept $\alpha(o)=1+\varepsilon >1$ (and then apply to it some a not well defined unitarization procedure) or we consider from the very beginning a more complicated pomeron singularity (for instance,  in the form (\ref{eq:tripole})). 
	The question may be asked: does froissaron satisfy the Dyson-Schwinger  equations (DSE),  provided that the 3-froissaron vertex is chosen appropriately	
	 
\begin{figure}[h]
		\begin{center}
			\includegraphics[scale=0.6]{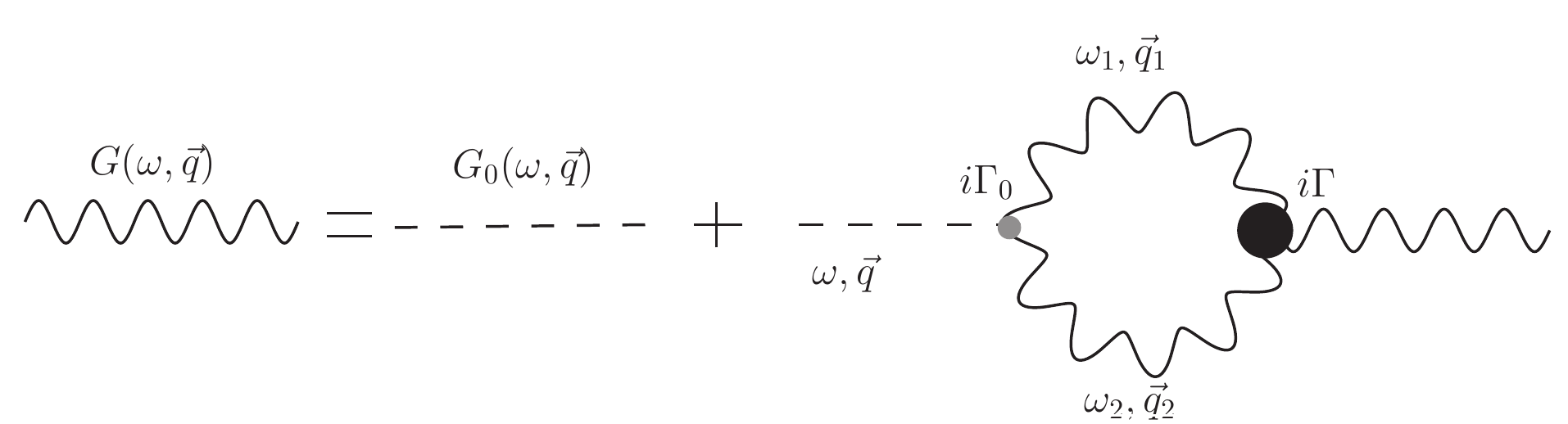}
			\caption{DSE for the pomeron propagator. Wave line is the full propagator, dashed line is the bare propagator. The small gray circle is the bare or input 3P-vertex, the black circle is the dressed or output vertex}
			\label{fig:DS-propagator}
		\end{center}
	\end{figure}

	The DS equation for propagator is given in the Fig. \ref{fig:DS-propagator}.
	
	Analytically DSE for propagator has the form:
	
	\begin{equation}\label{eq:propag-DS}
		G(\omega,\vec{q}\,^{2})=G_{0}(\omega,\vec{q}\,^{2})+ G_{0}(\omega,\vec{q}\,^{2})\Sigma (\omega,\vec{q}\,^{2})G(\omega,\vec{q}\,^{2})
	\end{equation}
	where
	\begin{equation}\label{eq:sigma-DS}
		\begin{aligned}
			\Sigma(\omega,\vec{q}\,^{2})&= -\displaystyle \frac{1}{2!}\int_{\uparrow}\frac{d\omega'}{2\pi i}\int\frac{d^{2}q'}{\pi}\Gamma_{0}(\{\omega\},\{q\})G(\omega ',\vec{q}\,'\,^{2})\\
			&\times G(\omega-\omega',(\vec{q}-\vec{q}\,')^{2})\Gamma(\{\omega\},\{q\}),
		\end{aligned}
	\end{equation}
	$\omega \equiv j - 1$, $j$ is the pomeron angle momentum, $\vec{q}$ is the transverse component of the pomeron momentum, $\vec{q}\,^2 \approx -t$.
	
\begin{figure}[h!]
		\begin{center}
			\includegraphics[scale=0.4]{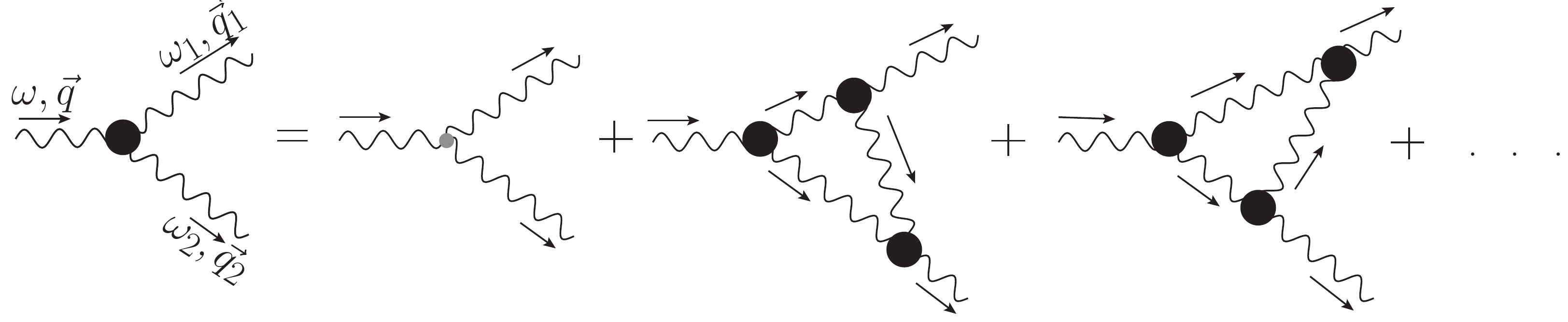}
			\caption{Dyson-Schwinger equation for 3P-vertex}
			\label{fig:DS-vertex}
		\end{center}
	\end{figure}

	The equation for 3-pomeron vertex $\Gamma(\omega,\omega_{1},\omega_{2};\vec {q},\vec {q}_{1},\vec {q}_{2})$ is given by the Fig. \ref{fig:DS-vertex}.
	
	In the leading order the analytical form of DSE for 3P-vertex (if only 3P-vertices are considered) is:
	
	\begin{equation}\label{eq:vertex-DS}
		\begin{aligned}
			\Gamma (\{\omega \};\{q\})&=\Gamma_{0} (\{\omega\};\{q\}) -\displaystyle \int\frac{d\omega'}{2\pi i}\int\frac{d^{2}q'}{\pi}\Gamma(\{\omega\}; \{q\})G(\omega',\vec{q}\,'^{2})\\&\\	
			&\times \Gamma(\{\omega\};\{q \}) G(\omega'-\omega_{1},(\vec{q}\,'-\vec{q}\,_{1})^{2})\Gamma(\{\omega \};\{q \})\\
			&\\
			&\times G(\omega-\omega',(\vec{q}-\vec{q}\,')^{2}))+(\omega_{1}\leftrightarrow \omega_{2},  \vec{q}_{1}\leftrightarrow \vec{q}_{2}),
		\end{aligned}
	\end{equation}	
	where the notations $\{\omega\}=\omega, \omega',\omega-\omega'$, $\{q\}=\vec{q}\,^{2},\vec{q}\,'^{2},(\vec{q}-\vec{q}\,')^{2}$ are used.

	General properties of DSE in the framework of Reggeon Field Theory (with the single pomeron  input pole)  were discussed in details by V.N. Gribov \cite{Gribov}. Two regimes, the weak coupling and the strong coupling ones, were analyzed. The weak coupling regime was noted as preferable, however it is not supported by available experimental data. 
	    
	The first attempt to discuss the froissaron (\ref{eq:tripole}) as input in DSE  was made by J.S.  Ball \cite{Ball}. He considered the model in which the  output propagator and 3f-vertex are proportional  to input ones
	\begin{equation}
		\begin{aligned}
		G(w,q^2)&=gG_0(\omega,q^2), \quad &\Gamma(\omega, \omega',q,q')&=\gamma		
\Gamma_0(\omega, \omega',q,q'),\\
G_0(\omega,q^2)&=2\pi a^2(\omega^2+a^2q^2)^{-3/2}, \quad &\Gamma_0(\omega, \omega',q,q')&=\gamma_0(\omega^2+a^2q^2)^{3/2}/(2\pi a^2)	
		\end{aligned} 
	\end{equation}
and have obtained two algebraic equations for couplings $g$ and $\gamma$.

However, one can see that in this simple scheme the integrated DSE cross section, $\sigma_{SDE}(s)=0$. We think that the problem can be fixed only if the input 3f-vertex is changed for a more complicated form. 

Developing this idea is one of the goals of our work. In Section \ref{sect: P&3F} we define the main ingredients in our scheme and estimate corrections to the input propagator and 3f-vertex in the DS equations. Differential and integrated cross sections of diffraction production processes are estimated at asymptotic energy in the Section \ref{sect:DiffCrSec}.

We would like to emphasize here that our approach is based on two main assumptions. \underline {The first:} input reggeon in DSE is a froissaron (Eq.\eqref{eq:tripole}), which for $ t = 0 $ is located in the $\omega$-plane at the point $ \omega = 0 $ not violating the Froissart-Martin bound. \underline{The second:} we assume that the 3f-vertex depends on the angular and spatial momenta of the froissarons in it, therefore the factorization of propagators and vertices takes place only in the ($ \omega, t $)-representation, but does not hold in the ($s,t$) one, which is valid for input pomeron in the form of a simple pole.

In the Section \ref{sect:DiffCrSec} cross sections of diffraction processes are calculated at $s\to \infty$. The limits for free parameters at which the diffraction cross sections do not exceed Froissart-Martin bound are obtained. The results are summarized in the Conclusion.

\section{Propagator and 3-F vertex. Restrictions on the vertex parameters}\label{sect: P&3F}

In accordance with a general form of the elastic scattering partial amplitude at low $\omega$ and $q_i^2$ we consider  the universal propagator for froissaron
\begin{equation}\label{eq:propagator-0}
	G_0(\omega,q)= \dfrac{E(\omega,q)}{(\omega^2+a^2q^2)^{3/2}} 
\end{equation} 
where function $E(\omega,q)$ is a finite function at any $\omega, \vec{q}$, providing the main contributions in the integrals over $\omega, \vec{q}$ in the region where $\omega^2\approx aq^2\approx 0$. We remind that our main interest is concentrated in the limit of high energies and low transferred momenta, which correspond to above mentioned $\omega\to 0$.

Now we suppose that in accordance with the structure of the Froissaron singularity in $G_0(\omega,q)$ at $\omega^2+\omega_{0}^2=0$ the function $E(\omega,q)$ depends on $\omega$ through the variable $\kappa =(\omega^2+\omega_{0}^2) ^{1/2}$  and it can be expanded in powers of $\kappa$:
\begin{equation}\label{eq:propagator-1}
	G_0(\omega,q)=\dfrac{ E_{0}(q)+\kappa E_{1}(q)+ \kappa^2 E_{2}(q)}{\kappa^{3}}=
	\sum\limits_{k=0}^{2} G_0^{(k)} (\omega,q) 
\end{equation}
where $k$ corresponds (at $q=0$) to the contribution of the triple pole ($k=0)$, double pole ($k=1$) and single pole ($k=2$).
Thus the frroissaron propagator can be written in the form with main and sub-asymptotic (SA corrections) terms
\begin{equation}\label{eq:F-propagator} 
	G_0(\omega,q)=
	\dfrac{E_{0}(q)}{(\omega^2+\omega_0^2)^{3/2}} + \dfrac{E_{1}(q)}{\omega^2+\omega_0^2}+
	\dfrac{E_{2}(q)} {(\omega^2+\omega_0^2)^{1/2}}, \quad \omega_0=aq.  
\end{equation}

Let us notice that the first terms in the Eq. \eqref{eq:F-propagator}  has a pair of branch points colliding at $\omega_0=0\,  (q=0)$ and generating a triple pole.  
The numerators $E_k(\omega,q, k=0,1,2)$ can be chosen for simplicity  in exponential form either $e^{-B_k q}$ or $e^{-B_k q^2}$, although it can be more sophisticated. The amplitudes with such terms can be calculated in $(s,b)$-representation, or at least can be estimated at $b\ll \xi=\ln(s/s_0)$ and $b\gg \xi=\ln(s/s_0)$. The details for $k=0$ are given in the  \ref{sect:Modified propagator}.

We impose certain requirements to the vertex function, from which the intervals for the vertex parameters can be set.

\begin{enumerate}
	\item Vertex could not have singularities in $\omega_i,q_i\sim 0$ which lead to its infinity and could not cancel a singularity of the propagator (while it can make it a more soft). It means that vertex can (must) have zeros at some of these variables.  
	\item  Diffractive differential and their integrated cross-sections could not violate the unitarity restriction;
	\item  In accordance with the experimental data differential cross section the single diffraction dissociation does not vanish at $t=0$;
	\item Most likely, the experiments show that diffractive integrated cross sections rise with energy slowly then total and elastic integrated cross sections.
	\item  Corrections to propagator and vertex in the DS-equations would be small at small $\omega_i$ and $q_i^2$;
\end{enumerate}

\begin{figure}[!h]
	\centering
	\includegraphics[width=0.4\linewidth]{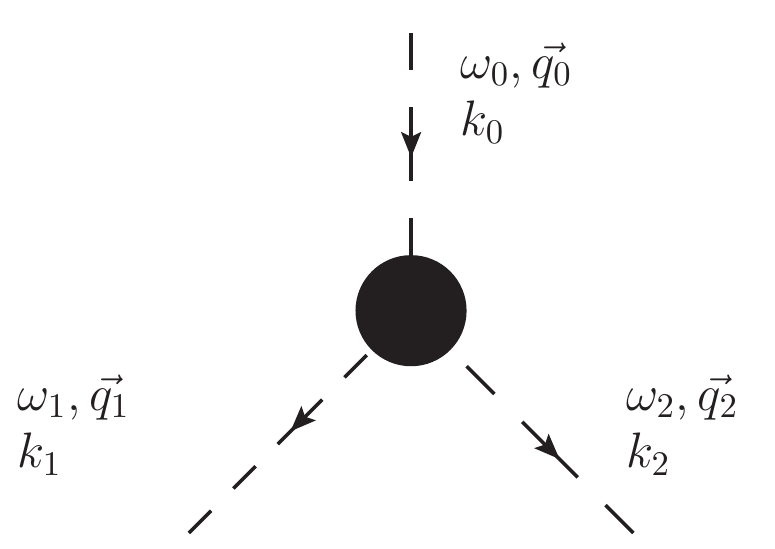}
	\caption{General 3-Reggeon (3R)  vertex depending on the kinds of legs}
	\label{fig:vertex-general}
\end{figure}

We consider here the 3f-vertex function in a factorized form. Each of the three factors  at the vertex  (Fig. \ref{fig:vertex-general}) depends on the type of reggeon corresponding to it in accordance with Eq. \eqref{eq:F-propagator} and has the same functional form.  Generally, the factor corresponding to the input froissaron (with $\omega_0, \vec{q_0}$) can  differ of other, output ones. For instance, it can have free parameters which are different from those at other factors shown in the next equations. We don't consider here such a possibility in order to avoid a non-principal complexity.

\begin{equation}\label{eq:3f-general}
	\begin{aligned}
		\Gamma_0 ^{(k,k_1,k_2)}(\omega,\omega_{1},\omega_{2},\vec{q},\vec{q}_{1},\vec{q}_{2})=
		\gamma_k(\kappa,q) \gamma_{k_1}(\kappa_{1},q_1) \gamma_{k_2}(\kappa_{2},q_2),\\
		\kappa_{i}=(\omega_i^2+\omega_{0i}^2) ^{1/2}=(\omega_i^2+a^2q_i^2) ^{1/2}, \quad \omega_i=\omega, \omega_1.\omega_2 
	\end{aligned}
\end{equation}
where   
\begin{equation}\label{eq:vert-factors}
	\gamma_i(\kappa_{i},q_i)=\gamma_i(0,q_i) \kappa_{i}^{\mu(k_i)} 
\end{equation}
and
\begin{equation}\label{eq:mu-k}
	\mu(k)=\mu_1(k)+k\\
\end{equation}
with $\mu_1(k)$ coming from the 3f-vertex while the second term $k$ takes into account the kind of reggeon from the expansion \eqref{eq:F-propagator} of the f-propagator.

To satisfy the point 1 from the list of requirements to 3f-vertex we must impose
\begin{equation}\label{eq:mu-limits}
	0<\mu_1(k)<3 \quad \text{at any} \quad k=0,1,2.
\end{equation}

It will be shown in the Section \ref{sec:SDD} that $\mu_1(k=0)\equiv \mu_0\neq 0$. Otherwise this term leads to the integrated diffraction cross section rising with energy $\propto \ln^5(s/s_0)$  (F-K problem). 

In the next Sections we consider three specified choices of the function $\mu_1(k)$ 
\begin{equation}\label{eq:vertex-variants} 
	\begin{aligned}
		&\rm{a)} \quad  \mu_1(k)=\mu_0,\\
		&\rm{b)} \quad 	\mu_1(k)=\mu_0(1+\lambda k),\\ 
		&\rm{c)} \quad 	\mu_1(k)=\dfrac{\mu_0}{1+\lambda k}. 
	\end{aligned}
\end{equation}

Variant a) chooses the universal form of 3f-vertex independent on the corrections to propagators. Variants b) and c) describe an increasing and decreasing  with $k$ power of $\kappa$ in the vertex.

Now let's consider a smallness of the corrections to propagator $G_0(\omega,t)$ and vertex  $\Gamma_0(\{\omega \}, \{\vec{q} \})$.  
\begin{figure}[!h]
	\centering
	\includegraphics[width=0.4\linewidth]{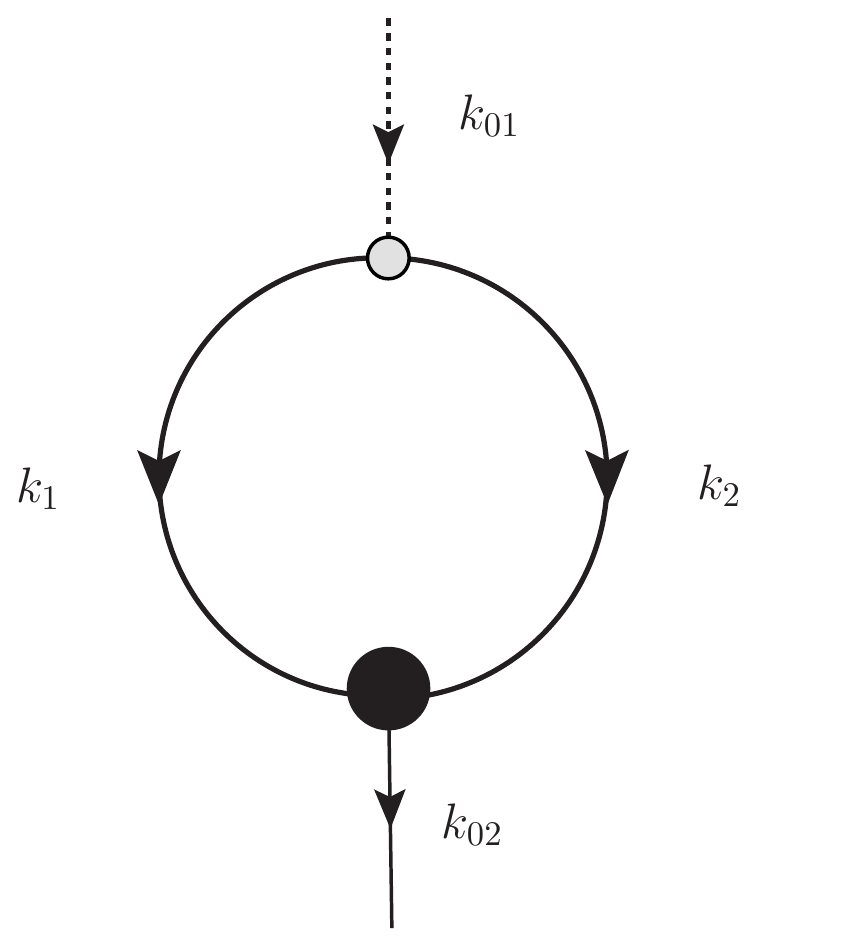}
	\caption{General form of the $\Sigma^{(k_{01},k_{02};k_1,k_2)}(\omega_0,q_0)$ diagram in the DS equation}
	\label{fig:loop-general}
\end{figure}

\subsection{DS-Corrections. All reggeons are froissarons}

Let's consider here the corrections of the kind 1. In the equations for propagator  \eqref{eq:propagator-1} and \eqref{eq:F-propagator} only the main term (with $k=0$) is taken into account. In this section we ignore the sub-asymptotic, coming from propagators 
corrections (Eq. \eqref{eq:propagator-1}) which are small in the considered here region of $\omega, q$.
\begin{equation}\label{eq:F-propagator-0} 
	G_0^{(0)}(\omega,q)=
	\dfrac{E_{0,+}(q)}{(\omega^2+\omega_0^2)^{3/2}} \quad \omega_0=\sqrt{a}q  
\end{equation}

\begin{equation}\label{eq:3f-vertex} 
	\begin{aligned}
		\Gamma_0 ^{(0)}(\omega,\omega_{1},\omega_{2},\vec{q},\vec{q}_{1},\vec{q}_{2})=
		\gamma_1(\kappa,q) \gamma_2(\kappa_{1},q_1) \gamma_3(\kappa_{2},q_2),\\
		\kappa_{i+}=(\omega_i^2+a^2_+q_i^2) ^{1/2}, \quad \omega_i=\omega, \omega_1.\omega_2 
	\end{aligned}
\end{equation}
where   
\begin{equation}\label{eq:vert-func}
	\gamma_i(\kappa_{i},q)=\gamma_i(0,q_i) \kappa_{i}^{\mu(k_i=0)}. 
\end{equation}

It will be shown in the Section \ref{sec:SDD} that $\mu(k=0)\equiv \mu_0\neq 0$. Otherwise this term leads to the integrated diffraction cross section rising with energy $\propto \ln^5(s/s_0)$   (this is the part of F-K problem or paradox).

\subsubsection{Smalness of the DS-corrections}\label{sect:DS-corrs}
\underline{\bf Propagator.}

Firstly, let's consider the ''simplest`` case with one parameter $\mu_0$ in the vertex $\Gamma$, all $k_i=0$. 
The input froissaron propagator $G_0^{(0)}$ and input vertex $\Gamma_0^{(0)}$ are defined by Eqs. \eqref{eq:F-propagator-0} and \eqref{eq:3f-vertex} correspondingly.

In this case we have for the first corrections to $G_0^{(0)}$-propagator (all propa\-gators and vertices in the first approximation are the input ones $G_0^{(0)}, \Gamma_0^{(0)}$):
\begin{equation}\label{eq:1st-G-corr}
	\begin{aligned}
		G_1(\omega,q)&=G_0^{(0)}(\omega,q^2)+\Delta G, \\
		\Delta G &= G_0^{(0)}(\omega,q)\Sigma_0^{(0)}(\omega,q)G_0^{(0)}(\omega,q),
	\end{aligned}
\end{equation}

\begin{equation}\label{eq:Sigloop}
	\begin{array}{ll}
		\Sigma_0^{{(0)}}(\omega,q)=&\\
		=\int\limits_C\dfrac{d\omega'}{2\pi i}\int\dfrac{d^2q'}{\pi}\Gamma_0^{(0)}(\{\omega\},\{ \vec{q} \}) G_0^{(0)}(\omega ,q_1) G_0^{(0)}(\omega,q_2)\Gamma_0^{(0)}(\{\omega\},\{\vec{q}\}).
	\end{array}
\end{equation}
\begin{figure}[!h]
	\centering
	\includegraphics[width=1.0\linewidth]{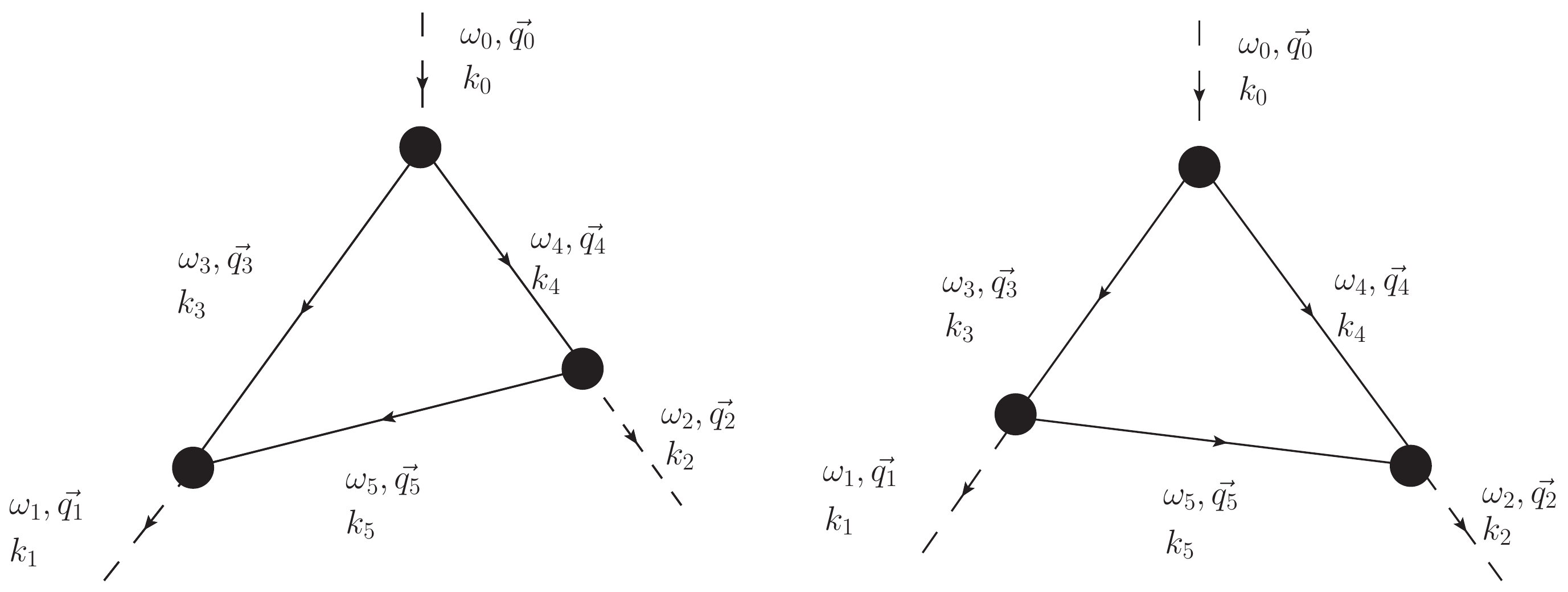}
	\caption{Corrections to 3-reggeon vertex}
	\label{fig:Gamma-corrs}
\end{figure}

Now let's us estimate the corrections to $G_0$ at $s\to \infty$.  In the Eq. \eqref{eq:Sigloop}  the essential region of integration is  $ \omega  (\omega')\sim aq (aq' )\sim 1/\xi $. 
Therefore, we have from Eqs. \eqref{eq:1st-G-corr}  and \eqref{eq:Sigloop}
\begin{equation}\label{eq:propagatot-ests}
	\Delta G\propto G_0^{(0)}(\omega^2)^{3/2} (\omega^2)^{(-3/2)}(\omega^2)^{3\mu_0}(\omega^2)^{-3}= G_0^{(0)}(\omega^2)^{3\mu_0-3},
\end{equation}	
i.e. 
\begin{equation}\label{eq:G-1}
	G_1=G_0^{(0)}[1+{\cal{O}}(\omega^{6(\mu_0-1)})].
\end{equation}
The first correction is small at $\omega\sim q\sim 0$ if  $\mu_0> 1$. It follows from this inequality and \eqref{eq:mu-limits} that
\begin{equation}\label{eq:mu0-1}
	1<\mu_0<3.
\end{equation}

\bigskip
 \underline{\bf Vertex.}

The first correction $\Delta_1\Gamma, \Delta_2\Gamma,$ (Fig. \ref{fig:Gamma-corrs}) 
are the following 
\begin{equation}\label{eq:Sig-loop-0}
	\begin{aligned}
		&\Gamma_1(\omega_0,\omega_1,\omega_2, \vec {q_0}, \vec {q_1}, \vec {q_2})\equiv \Gamma_1 =\Gamma_0^{(0)} +\Delta_1\Gamma +\Delta_2\Gamma,\\
		&\Delta_1\Gamma= \int \dfrac{d\omega' d^2\vec{q'}}{2\pi i} \Gamma_0^{(0)}(\omega_0,\omega_3,\omega_4, \vec {q_0}, \vec {q_3}, \vec {q_4})\Gamma_0^{(0)}(\omega_1,\omega_3,\omega_5, \vec {q_1}, \vec {q_3}, \vec {q_5})\\
		&\times \Gamma_0^{(0)} (\omega_3,\omega_5,\omega_2, \vec {q_3}, \vec {q_5}, \vec {q_2})G_0^{(0)}(\omega_3,q_3^2)G_0^{(0)}(\omega_4,q_4^2)G_0^{(0)}(\omega_5,q_5^2),\\ 
		&\Delta_2\Gamma=\Delta_1\Gamma(\omega_1\leftrightarrow \omega_2).
	\end{aligned}
\end{equation}

Similarly to the propagator case  consideration of the correction to $\Gamma_0^{(0)}$ leads to
\begin{equation}\label{eq:vertex-ests}
	\Delta_1\Gamma\propto \Gamma_0^{(0)}(\omega^2)^{3/2}(\omega^2)^{6\cdot \mu_0/2} (\omega^2)^{-9/2}=\Gamma_0^{(0)}(\omega^2)^{3(\mu_0 -1)}.	
\end{equation}
Because $\Delta_1\Gamma \approx \Delta_2\Gamma $, we have in the considered limit 
\begin{equation}\label{eq:Gamma-1}
	\Gamma_1=\Gamma_0^{(0)}[1+{\cal{O}}\left (\omega^{6(\mu_0-1)}\right )].
\end{equation}
Again, correction is small at $\omega\sim q\sim 0$ if  $\mu_0> 1$.

Moreover, let's evaluate the  factor coming from one additional reggeon line between any two reggeon lines. We have 3 new propagators, 2 new 3f-vertices and one integral over new loop. The additional factor has at $\omega_i\sim ,q_i<<1$ the following behavior 
\begin{equation}\label{eq:+one-line}
	(\omega^2)^{3/2}(\omega^2)^{-9/2}(\omega^2)^{(6\mu_0)/2}=\omega^{6(\mu_0-1)}.	
\end{equation} 
Inserting   one reggeon loop into reggeon line we have the  same additional factor $\omega^{12(\mu_0-1)}$. And finally, inserting one reggeon line with one loop (6 new reggeons, 4 new vertices 2 new loops), we again obtain the same factor.  

This means that an increase in the number of reggeons and vertices only increases the number of ever-smaller corrections in DS-equations if $\mu_0>1$.

\subsection{DS-Corrections. General case}

Here we consider the general term defined in the propagator Eqs. \eqref{eq:propagator-1} and \eqref{eq:F-propagator}
\begin{equation}\label{eq:propagator-general}
	G_0^{(k)}(\omega_k,q_k)=
	\dfrac{E_{k}(q_k)}{(\omega_l^2+\omega_{0l}^2)^{3/2-k/2}}, \quad \omega_{0l}=aq_k, \quad k=0,1,2. 
\end{equation}
and in  3R-vertices
\begin{equation}\label{eq:3R-general}
	\begin{aligned}
		\Gamma_0 ^{(k_0,k_1,k_2)}(\omega,\omega_{1},\omega_{2},\vec{q},\vec{q}_{1},\vec{q}_{2})=
		\gamma_{k_0}(\kappa,q) \gamma_{k_1}(\kappa_{1},q_1) \gamma_{k_2}(\kappa_{2},q_2^2),\\
		\kappa_{i}=(\omega_i^2+\omega_{0i}^2) ^{1/2}=(\omega_i^2+a^2q_i^2) ^{1/2}, \quad \omega_i=\omega, \omega_1.\omega_2 
	\end{aligned}
\end{equation}
It should be noted that $k_1, k_2$ values must be equal to those in the  corresponding terms of the left and the right reggeon's propagators in the given 3f-vertex.  
Functions $\gamma(\omega,k)$ in the Eq. \eqref{eq:3R-general} we choose in the form
\begin{equation}\label{eq:mu_i-3P-vertex}
	\gamma_i(\omega_i,\kappa_{i})=\gamma_i(0,q_i) \kappa_{i}^{\mu(k_i)}.
\end{equation}

The Eqs. \eqref{eq:1st-G-corr}, \eqref{eq:Sigloop} in the case $k_i\neq 0$ are transformed to
\begin{equation}\label{eq:1st-G-corr-k}
	\begin{aligned}
		G_1^{(k)} (\omega,q)&
		\stackrel{\omega, q\to 0}{\approx }
		G_0^{(k)}(\omega,q) \left [1+\Delta G^{(l} \right ], \\
		\Delta G^{(k)} &=
		\sum\limits_{k,k_1,k_2=0}^{2}G_0^{(k)}(\omega,q_l) \Sigma_0^{(k,k_1,k_2)}(\omega,q_l).
	\end{aligned}
\end{equation}
Now we can take into account that $G_0^{(k=0)} >>G_0^{(k=1)}>>G_0^{(k=2)}$ in the considered region of $\omega, q$. (For $k\neq 0$ we have in $\Delta G^{(l)}$ additional small factors $(\omega^2)^n$ where $n>0$ and it depends on the values of $k,k_1,k_2$.)  Thus, we come back to the results \eqref{eq:G-1}, \eqref{eq:mu0-1}, \eqref {eq:Gamma-1}. 

 It is found in the Section \ref{sec:CDP} additional inequality 
\begin{equation}\label{eq:ineq-mu-0}
	1<\mu_0<3/2.
\end{equation}     
For  corrections to the vertex function one can obtain the following estimations
\begin{equation} \label{eq:1st-Gamma-corr-k}
	\begin{aligned}
		\Gamma_1^{{(k,k_1,k_2)}}&\propto \Gamma_0^{{(k,k_1,k_2)}}[1+\Delta \Gamma^{(k)}],\\
		\Delta \Gamma^{(k)}&\propto \sum\limits_{k_3,k_4,k_5=0}^2 (\omega^2)^{-3+\mu(k_3)+\mu(k_4)+\mu(k_5)+k_3+k_4+k_5}\\
		=&\sum\limits_{k_3,k_4,k_5=0}^2(\omega^2)^{3(\mu_0-1)+\mu(k,k_1,k_2)},
   \end{aligned}
\end{equation}
where
\begin{equation}\label{eq:mu-3k}
	\mu(k,k_1,k_2)=\sum\limits_{i=3}^5(\mu(k_i)-\mu_0+k_i).
\end{equation}
Function $\mu(k,k_1,k_2)$ is positive for any values of $0 \leq k_i \leq 2$ and equal to zero if $k_i=0$. It is valid for both choices of $\mu_1(k)$ in Eq. \eqref{eq:vertex-variants}. Thus, in the  general case  a more small corrections comparing with the main ones (at all $k=0$) are added. 

In the \ref{sect:limits-params} the inequalities for parameters $\mu_0, \lambda$ in three choices for the function $\mu_1(k)$ defined in the Eq.  \eqref{eq:vertex-variants} are obtained.
Combining all these constraints  we get 
\begin{equation}\label{eq:ineqs-mu-0-lambda}
	\begin{aligned}
1<&\mu_0<3/2,\\
\quad &\lambda=0 &\rm for\,\, the\,\ case\,\, a),\\
-1/2<&\lambda<-/1/4  \quad &\rm for\,\, the\,\ case\,\, b),\\
&\lambda>0  \quad  &\rm for\,\, the\,\ case\,\,  c).
	\end{aligned}
\end{equation}

\section {Diffraction processes with large rapidity gaps}\label{sect:DiffCrSec}

\medskip
\subsection{Single diffraction dissociation (SDD)}\label{sec:SDD}

\begin{figure}[h]
	\begin{center}
		\includegraphics[scale=0.5]{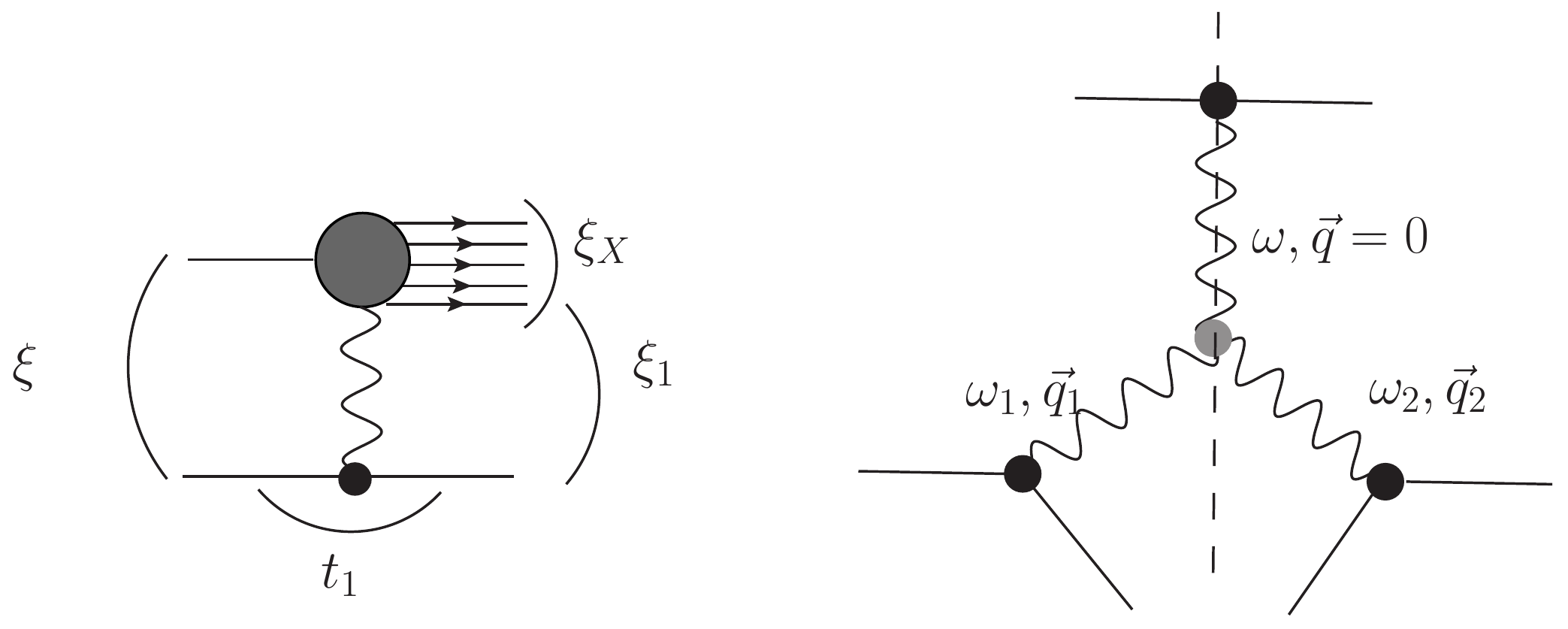}
		\caption{SDD process (left) and the corresponding diagram (right) from generalized optical theorem, $\xi=\xi_{X}+\xi_{1}$}
		\label{fig:SDD}
	\end{center}
\end{figure}

In the general case the  input propagator $G$ and $3f$-vertex with the parameters ($\mu_0, \lambda$) are defined by Eqs. \eqref{eq:propagator-general}, \eqref{eq:3R-general}, \eqref{eq:mu_i-3P-vertex}.
\begin{equation}\label{eq:SDD-k-1}
	\sigma_{SDD}(s)=\sum\limits_{k_0,k_1,k_2=0}^2\sigma_{SDD}^{(k_0,k_1,k_2)}(s)
\end{equation}
where
\begin{equation}\label{eq:SDD-k-2}
	\sigma_{SDD}^{(k_0,k_1,k_2)}(s)=\displaystyle \int\limits_{\xi_0}^{\xi-\xi_0 }d\xi_{X}\int\limits_{0}^{\infty}dq\,q\dfrac{d\sigma_{SDD}^{(k_0,k_1,k_2)}}{dtd\xi_X}
\end{equation}
and  
\begin{equation}
	\begin{aligned}
		\displaystyle \dfrac{d\sigma_{SDD}^{(k_0,k_1,k_2)}} {dtd\xi_{X}}&=\displaystyle C_{SDD}   \int\frac{d\omega }{2\pi i}\int\frac{d\omega_{1}}{2\pi i} \int\dfrac{d\omega_{2}}{2\pi i}e^{\xi_{X}\omega }e^{\xi_{1}(\omega_{1}+\omega_{2})}\\
		&\times  \eta_{\omega_{1}}\eta^{*}_{\omega_{2}}v_{k_1}(t)v_{k_2}(t)E_{k_0}(0)E_{k_1}(t)E_{k_2}(t) \dfrac{\kappa_{0}^{\mu_1({k_{0})}}\kappa_{1}^{\mu_1({k_{1})}}\kappa_{2}^{\mu_1({k_{2})}}}{\kappa_{0}^{3-k_0}\kappa_{1}^{3-k_1}\kappa_{2}^{3-k_2}}\\
		&=\tilde C_{SDD} 
		\displaystyle \int\dfrac{d\omega }{2\pi i} \dfrac{e^{\xi_{X}\omega }}{\omega^{3-\mu(k_{0})}}
		\int \dfrac{d\omega_1 }{2\pi i} 
		\dfrac {e^{\xi_1\omega_1}}{\kappa_{1}^{3-\mu(k_{1})}}
		\int\dfrac{d\omega_2}{2\pi i} \dfrac{e^{\xi_1\omega_{2}}}{\kappa_{2}^{3-\mu(k_{2})}}\\
		&\times \eta_{\omega_1}\eta^{*}_{\omega_2}v^2(t)E_{k_0}(0)E_{k_1}(t)E_{k_2}(t)  
	\end{aligned}
\end{equation}
where
\begin{equation}\label{eq:def-C-mu}
	\begin{aligned}
		C_{SDD}&=\frac{1}{32\pi^{2}}v_{k_0}(0)\gamma_{k_0}(0,0)\gamma_{k_1}(0,0)\gamma_{k_1}(0,0),\\ 
		E_{k_{i}}(t)&=\exp(B_{k_{i}}t). \quad v_{k_i}(t)=v_{k_i}(0)\exp(B_{v,k_i}vt)\\
    	\mu(k)&=\mu_1(k)+k.
	\end{aligned}
\end{equation}
and $\mu_1(k)$ is defined in Eq. \eqref{eq:vertex-variants}.

Then, after integration over $\omega$ the  differential SDD cross section can be written in the following form
\begin{equation}\label{eq:int-omega-gen}
	\begin{aligned}
		\dfrac{d\sigma_{SDD}^{(k_0,k_1,k_2)}}{dtd\xi_{X}}&= \tilde C_{SDD} 	
		 \displaystyle \left (\dfrac{J_{1-\mu(k_1)/2}(\tilde \xi_{1}aq)}{\Gamma ((3-\mu(k_1)/2)}\left (\dfrac{\tilde \xi_{1}}{2aq}\right )^{1-\mu(k_1)/2}\right )E_{k_1}(q^2)v_{k_1}(q^2)
		\\		&\times 
		\left (\dfrac{J_{1-\mu(k_2)/2}(\tilde \xi_{1}^*aq)}{\Gamma ((3-\mu(k_2)/2)}\left (\dfrac{\tilde \xi_{1}^*}{2aq}\right )^{1-\mu(k_2)/2}\right )	E_{k_2}(q^2)v_{k_2}(q^2),\\
		\tilde C_{SDD} &=C_{SDD} \dfrac{2\pi 2^{-2+\mu(k_1)/2+\mu(k_2)/2 }} 
{\Gamma  (3-\mu(k_0))\Gamma((3-\mu(k_1))/2)\Gamma((3-\mu(k_2))/2)}
	\end{aligned}
\end{equation}
where $\tilde \xi_{1}=\xi_{1}-i\pi/2,\, \xi_1=\xi-\xi_{X}$. The given presentations ot integrals over $\omega_i$ are valid if $3-\mu_1(k_i)>0$. 

Restrictions on the parameters $\mu_0, \lambda $ in function $\mu_1(k)$ have been derived in \ref{sect:limits-params}.

\begin{equation}\label{eq:SDD}
	\begin{array}{ll}
		\sigma^{(k_0,k_1,k_2)}_{SDD}(s)&\approx \tilde C_{SDD} \displaystyle \sum\limits_{k_0,k_1,k_2=0}^{2}\int\limits_{\xi_0}^{\xi-\xi_0 }d\xi_{X}\int\limits_{0}^{\infty}dq\,q\frac{d\sigma^{(k_0,k_1,k_2)}_{SDD}}{dtd\xi_{X}}\\
		& \approx \tilde C_{SDD}\displaystyle \sum\limits_{k,k_1,k_2=0}^{2}
		 \displaystyle  v_{k_1}v_{k_2}  \int\limits_{\xi_0}^{\xi-\xi_0  }d\xi_{X}\xi_{X}^{2-\mu(k_0)}(\xi-\xi_{X})^{2-\mu(k_1)-\mu(k_2)}\\
		&\times \displaystyle \int\limits_{0}^{\infty}dx \,x\left (\dfrac{J_{1-\mu(k_1)/2}(\tilde x)}{\tilde  x^{1-\mu(k_1)/2}}\right )\left (\dfrac{J_{1-\mu(k_2)/2}(\tilde x)}{\tilde x^{1-\mu(k_2)/2}}\right )^*\\
		&\propto  \displaystyle \sum\limits_{k_0,k_1,k_2=0}^{2}\int\limits_{\xi_0}^{\xi-\xi_0  }d\xi_{X}\xi_{X}^{2-\mu(k_0)}(\xi-\xi_{X})^{2-\mu(k_1)-\mu(k_2)}
	\end{array}
\end{equation}
where we put $E_{k_i}(q) \approx 1, v_{k_i } (q) \approx v_{k_i}(0)\equiv v_{k_i}$

Cross section energy dependence is governed by the integral
\begin{equation}\nonumber
	\begin{array}{ll}
		&  \int\limits_{\xi_0}^{\xi-\xi_0 }d\xi_{X}\xi_{X}^{2-\mu(k_0)-k_0}(\xi-\xi_{X})^{2-\mu(k_1)-\mu(k_2)-k_1-k_2}=\xi^{5-\mu(k_0)-\mu(k_1)-\mu(k_2)-(k_0+k_1+k_2)}\\
		&\times \int\limits_{\xi_0}^{1-\xi_0/\xi}dx  x^{2-\mu(k_0)-k_0}(1-x)^{2-\mu(k_1)-\mu(k_21)-k_1-k_2}
		=\xi^{-1+\alpha_1+\alpha_2} I_{SDD}(\xi) 
	\end{array}
\end{equation}
where $\alpha_1=3-\mu(k_0)-k_0>0, \quad \alpha_2=3-\mu(k_1)-\mu(k_2)-k_1-k_2$ (sign of $\alpha_2$ depends on the values of $k_i$).

The estimation of the $I_{SDD}(\xi)$ is given in the  \ref{sect:SDD-cs}. So, the final result is the following
\begin{equation}
	\begin{aligned}
		\sigma_{SDD}^{(k_0,k_1,k_2)}(s)&\propto 
		\displaystyle \int\limits_{0}^{\xi}d\xi_{X}\xi_X^{2-\mu_1(k_0)} (\xi-\xi_X)^{2-\mu_1(k_1)-\mu_1(k_2)}\\
		& \propto 
		\xi^{2-3(\mu_0-1)}\times
		\left \{
		\begin{aligned}
			b) &\quad  \xi^{\lambda \mu_0 (k_0+k_1+k_2)},\,\, &-1/2<\lambda <0,\\
			c)  &\quad \xi^{-\lambda \mu_0\sum\limits_{i=0}^{2}\frac{k_i}{1+\lambda k_i}}, &\lambda >0.
		\end{aligned}  
		\right .
	\end{aligned}
\end{equation}

Thus, if $\mu_0>1$, then integrated cross section $\sigma_{SDD}^{(k_0,k_1,k_2)}(s) $  rises with energy slower than $\xi^2$ at any $0\leq k_0,k_1,k_2\leq 2$.

At $s\to \infty$ the main  contribution (it has  all $k_i=0$) to SDD cross section is
\begin{equation}\label{eq:SDD-main}
\sigma_{SDD}(s)\approx \sigma_{SDD}^{(0,0,0)}(s)\propto \xi^{2-3(\mu_0-1)}.	
\end{equation}

\medskip
\subsection{Central Diffraction Production (CDP)}\label{sec:CDP}

\begin{figure}[!h]
	\begin{center}
		\includegraphics[scale=0.6]{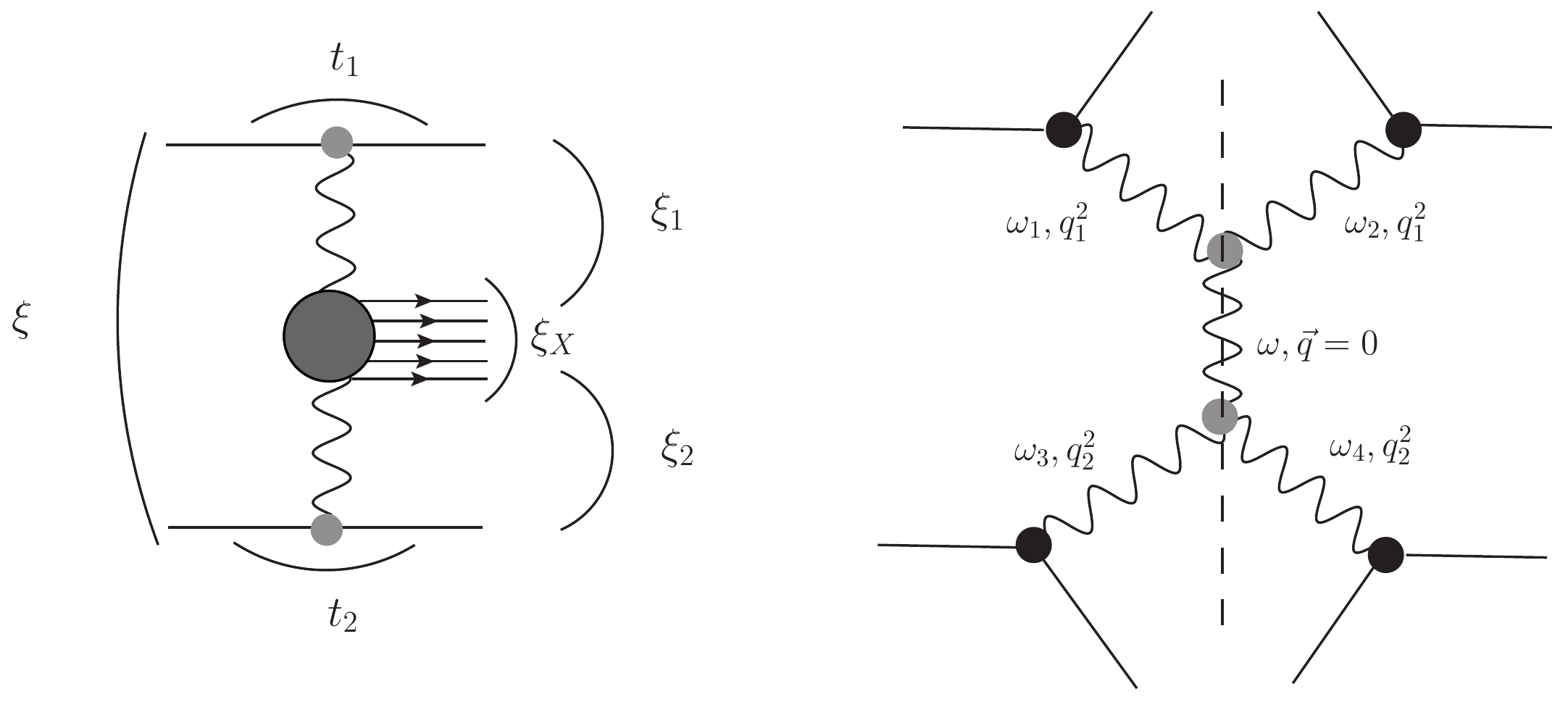} 
		\caption{CDP process (left) and the corresponding diagram (right) from generalized optical theorem, $\xi =\xi _{X}+\xi _{1}+\xi _{2}$}
		\label{fig:CDP-2}
	\end{center}
\end{figure}

Here and in the following subsections we consider only the main contributions to diffraction cross sections, because similarly to SSD case other terms at $s\to \infty$ rise  more slowly.

Let us write the expression for differential CDP cross section in terms of the above proposed propagators and vertices at all $k_i=0$:
\begin{equation} \label{eq:difCDP}
	\begin{array}{ll}
		\dfrac{d\sigma^{CDP}(\xi )}{dt_{1}d\xi_{1}dt_{2}d\xi_{2}}&=  C_{CDP}
		\displaystyle	\int\dfrac{d\omega }{2\pi i} \int\dfrac{d\omega_{1}}{2\pi i}
		\int\dfrac{d\omega_{2}}{2\pi i}\int\dfrac{d\omega_{3}}{2\pi i}\int\dfrac{d\omega_{4}}{2\pi i}\\
		&\times \eta_{\omega_{1}}\eta^{*}_{\omega_{2}}\eta_{\omega_{3}} \eta^{*}_{\omega_{4}}e^{\xi_{X}\omega}e^{\xi_{1}(\omega_{1}+\omega_{2})}e^{\xi_{2}(\omega_{3}+\omega_{4})}
		\dfrac{(\kappa^{2}\kappa_{1+}\kappa_{2+}\kappa_{3+}\kappa_{4+})^{\mu_0}} {\kappa^{3}\kappa_{1+}^{3}\kappa_{2+}^{3}\kappa_{3+}^{3}\kappa_{4+}^{3}}\\    
		&= C_{CDP} 
		\displaystyle \int\dfrac{d\omega }{2\pi i} \dfrac{e^{\xi_{X}\omega }}{\omega^{3-2\mu_0}}
		\int \dfrac{d\omega_1 }{2\pi i}\dfrac {\eta_{\omega_{1}}e^{\xi_1\omega_1}}{(\omega_1^2+\omega_{0,1}^2)^{(3-\mu_0)/2}}\\
		&\times \displaystyle \int\dfrac{d\omega_2}{2\pi i} \dfrac{\eta^{*}_{\omega_{2}}e^{\xi_1\omega_{2}}}{(\omega_2^2+\omega_{0,1}^2)^{(3-\mu_0)/2}}\\
		&\times \displaystyle \int \dfrac{d\omega_3 }{2\pi i}\dfrac {\eta_{\omega_{3}}e^{\xi_2\omega_3}}{(\omega_3^2+\omega_{0,2}^2)^{(3-\mu_0)/2}}
		\int\dfrac{d\omega_4}{2\pi i} \dfrac{\eta^{*}_{\omega_{42}}e^{\xi_2\omega_{4}}}{(\omega_4^2+\omega_{0,2}^2)^{(3-\mu_0)/2}}\\
	\end{array}
\end{equation}
where 
\begin{equation}
	\begin{aligned}
 C_{CDP}&=\frac{1}{32\pi^{2}}v_0(0,0)v^2_0(t,0)\gamma^2_0(0,0)\gamma^2_0(0,q_1^2)\gamma^2_0(0,q_2^2)E^2(t_1)E^2(t_2),\\ 
		E(t_1)&=\exp(t_1B_1).\,\, E(t_2)=\exp(t_2B_2).
	\end{aligned}
\end{equation}
Integral over $\omega$ in \eqref{eq:difCDP} is converged if $\mu_0<3/2$. 

\begin{equation} \nonumber
	\begin{aligned} 
		&	\dfrac{d\sigma^{CDP}}{dt_{1}d\xi_{1}dt_{2}d\xi_{2}}\propto \xi_X^{2-2\mu_0} \\
		&\times 	\left (\dfrac{J_{1-\mu_0/2}(\tilde \xi_{1}aq_1)}{\Gamma ((3-\mu_0)/2)}\left (\dfrac{\tilde \xi_{1}}{2aq}\right )^{1-\mu_0/2}\right )
		\left (\dfrac{J_{1-\mu_0/2}(\tilde \xi_{1}^*aq_1)}{\Gamma ((3-\mu_0)/2)}\left (\dfrac{\tilde \xi_{1}^*}{2aq}\right )^{1-\mu_0/2}\right )\\			
		&\times 
		\left (\dfrac{J_{1-\mu_0/2}(\tilde \xi_{2}aq_2)}{\Gamma ((3-\mu_0)/2)}\left (\dfrac{\tilde \xi_{2}}{2aq}\right )^{1-\mu_0/2}\right )
		\left (\dfrac{J_{1-\mu_0/2}(\tilde \xi_{2}^*aq_1)}{\Gamma ((3-\mu_0)/2)}\left (\dfrac{\tilde \xi_{2}^*}{2aq}\right )^{1-\mu_0/2}\right ).	\\
	\end{aligned}
\end{equation}
Similarly to the SDD case, we can estimate CDP cross section as
\begin{equation}\label{eq:int CDP}
	\begin{array}{ll}
		\sigma_{CDP}(s)&\propto \int\limits_{0}^{\xi} d\xi_1 \xi_1^{2-2\mu_0} \int\limits_{0}^{\xi-\xi_1} d\xi_2 \xi_2^{2-2\mu_0}(\xi-\xi_1-\xi_2)^{2-2\mu_0}\\
		&\\ 
		&\approx\xi^{8-6\mu_0}\int\limits_{0}^{1} dx_1x_1^{2-\mu_0} \int\limits_{0}^{1-x_1} dx_2 x_2^{2-\mu_0}(1-x_1-x_2)^{2-2\mu_0}\\
	\end{array}
\end{equation}
where $\xi_X=\xi-\xi_1-\xi_2$.  The intgrated CDP cross section rises slower than $\xi^2 $ if $8-6\mu_0<2$. Hence, 
\begin{equation}\label{eq:m_0CDP}
	1<\mu_0<3/2.
\end{equation}

\medskip
\subsubsection{Double diffraction dissociation (DDD)}

Let us consider the production of two bunches $X_1, X_2$ of rapidity widths $\xi_{X_{1}}$ $\xi_{X_{2}}$, separated by large rapidity gap $\xi_{1}=\xi-\xi_{X_{1}}-\xi_{X_{2}}$
\begin{figure}[h]
	\begin{center}
		\includegraphics[scale=0.5]{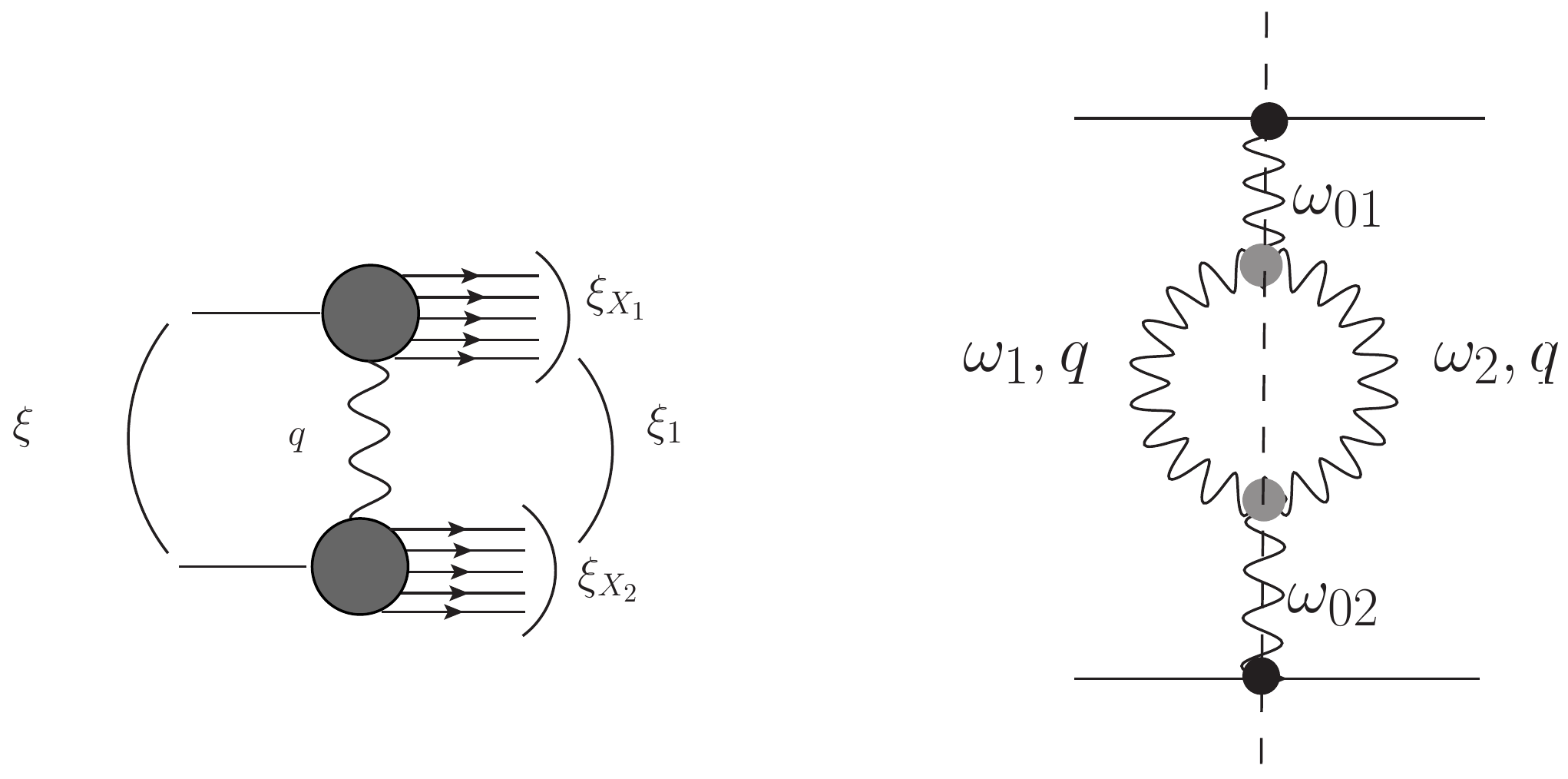}
		\caption{DDD process (left) and the corresponding diagram (right) from generalized optical theorem, $\xi=\xi_{X1}+\xi_{X2}+\xi_{1}$}
		\label{DDD-f}
	\end{center}
\end{figure}
\begin{equation}
	\begin{array}{ll}
	\displaystyle \frac{d\sigma_{DDD}}{dtd\xi_{1X}d\xi_{2X}} 
	& =\displaystyle C_{DDD}\int\frac{d\omega_{01}}{2\pi i}\int\frac{d\omega_{02}}{2\pi i}\int\frac{d\omega_{1}}{2\pi i}\int\frac{d\omega_{2}}{2\pi i}\\
	&\times\eta_{\omega_{1}}\eta^{*}_{\omega_{2}}e^{\xi_{1X}\omega_{01}}e^{\xi_{2X}\omega_{02}}e^{\xi_1(\omega_{1}+\omega_{2})} \dfrac{(k_{01}k_{02}k^2_{1+}k^2_{2+}))^{\mu_0}} {\kappa_{01}^3\kappa_{02}^{3}\kappa_{1+}^{3}\kappa_{2+}^{3}} \\
	&=C_{DDD}\displaystyle \dfrac{(\xi_{X1}\xi_{X2})^{2-\mu_0}}{\Gamma^2(3-\mu_0)} 
	\displaystyle \int\frac{d\omega_{1}}{2\pi}\int\frac{d\omega_{2}}{2\pi i}\\
	&\times \dfrac {e^{ \xi'_{1}\omega_{1}+ \xi_{1}^{'*}\omega_{2}  }}{(\omega_{1}^{2}+a^{2}q^{2})^{3/2-\mu_0 }(\omega_{2}^{2}+a^{2}q^{2})^{3/2-\mu_0}}\\ 
	&\propto (\xi_{X1}\xi_{X2})^{2-\mu_0}
	\displaystyle \left | \left (\dfrac{\xi'_{1}}{2aq}\right )^{1-\mu_0} J_{1-\mu_0 }(\xi'_{1}aq)\right |^{2}
\end{array}
\end{equation}
where $\xi'_1=\xi-i\pi/2-\xi_{1X}-\xi_{2X}$.

\begin{equation}\label{eq:DDD-1}
\begin{aligned}
	&\sigma_{DDD}\propto \displaystyle \int\limits_{\xi_0}^{\xi-2\xi_0}d\xi_{X1} \int\limits_{\xi_0}^{\xi-\xi_0-\xi_{1}} d\xi_{X2}\xi_{X1}^{2-\mu_0 }\xi_{X2}^{2-\mu_0 } \int\limits_0^{\infty} dq  q \left |{ \left (\dfrac{\tilde\xi_{1}}{2aq}\right )^{1-\mu_0 }}J_{1-\mu_0 }(\tilde \xi_{1}aq)\right | ^{2}\\
	&=\displaystyle \int\limits_{\xi_0}^{\xi-2\delta} d\xi_{X1} \int \limits_{\xi_0}^{\xi-\xi_0-\xi_{1}} d\xi_{X2}
	\xi_{X1}^{2-\mu_0} \xi_{X2}^{2-\mu_0 }|(\xi-\xi_{X1}-\xi_{X2})|^{2-4\mu_0}\int \limits_0^{\infty} dz z \left |\frac{\tilde J_{1-\mu_0}(\tilde z)}{z^{1-\mu_0}}\right |^{2}\\
	&\propto \xi^{8-6\mu_0} \displaystyle \int\limits_\delta^{1-2\delta}dx_1x_1^{2-\mu_0 } \int\limits_\delta^{1-\delta-x_1} dx_2 x_2^{2-\mu_0}(1-x_1-x_2)^{2 -4\mu_0}
	\propto \xi^2\xi^{-6(\mu_0-1)}.
\end{aligned}
\end{equation}
Because of  unitarity restriction on $\sigma_{DDD}(s)$ at $s\to \infty$ we  demand  $\mu_0>1$.

\medskip
\subsubsection{DDD with additional showers}

\begin{figure}[hbt!]
	\centering
	\includegraphics[width=0.45\linewidth]{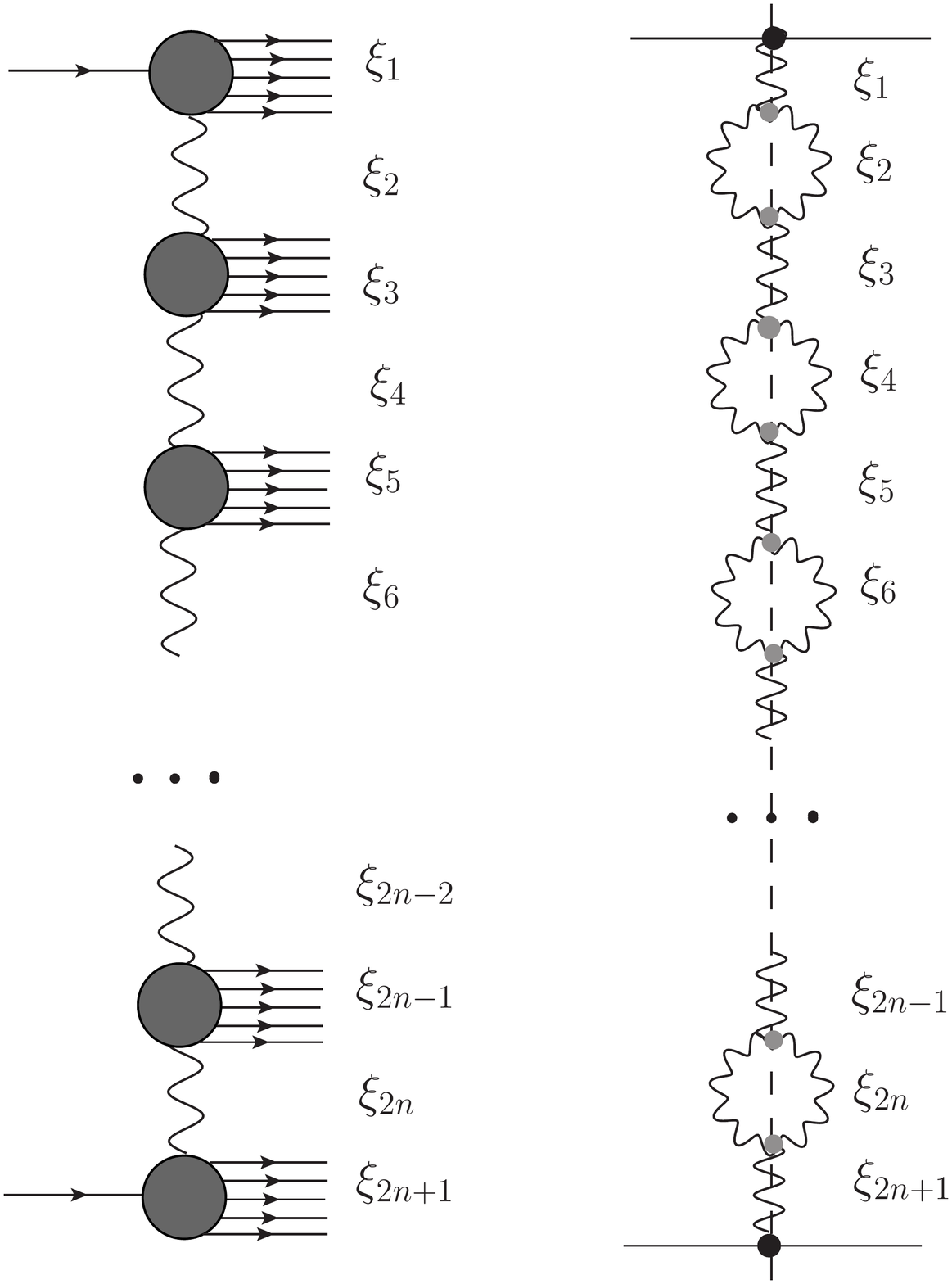}
	\caption{DDD with additional $n-1$ ''internal`` hadron showers}
	\label{fig:ddd-n}
\end{figure}
Now we can generalize DDD for a production of  more then two hadron showers separated by large rapidity gaps (the left diagram of fig. \ref{fig:ddd-n}). The right diagram comes from the generlized optical theorem for the process amplitude.

To estimate contribution of the given process to $\sigma_{tot}(s)$, we can write it in the following form keeping only the dependence of $\xi_i$. It can be made easily by evident extension of the Eq. (\ref{eq:DDD-1}) to arbitrary $n\geq 1$. However, it is evident that maximal value of $n$ is  $n_{max}=[(\xi/\xi_0-1)/2]$. Here we show the result for the main contribution (all $k_i=0t$ to $\sigma_{DDD}^{(n)}(s)$.

\begin{equation}\label{eq:DDD-n0}
	\begin{array}{ll}
		\sigma_{DDD}^{(n)}(s)&\propto \int\limits_{\Xi}\prod\limits_{i=1}^{2n+1}\!\!d\xi_i\,\,\delta \left (\xi-\sum\limits_{i=1}^{2n+1}\xi_i\right )  \xi_1^{2-\mu_0}\xi_{2n+1}^{2-\mu_0}\prod\limits_{i=1}^n\xi_{2i}^{2-4\mu_0}\prod\limits_{i=1}^{n-1}\xi_{2i+1}^{2-2\mu_0}\\
		\\
		&\propto \xi^{2-6n(\mu_0-1)},\qquad \qquad \sigma_{DDD}^{(1)}(s)\equiv \sigma_{DDD}(s).
	\end{array}
\end{equation}
Again, there is no a violation of $s$-channel unitarity if $\mu_0>1$. However we remind that corrections to froissaron propagator and 3f-vertex in DS equations are small at small $\omega$ and $q$ if $\mu_0>1$

Similarly one can obtain for other generalized processes
\begin{equation}\label{eq:D-n}
	\begin{aligned}
		\sigma_{SDD}^{(n)}(s)&\propto \xi^{2-3(1+2n)(\mu_0-1)}, \qquad \sigma_{SDD}^{(0)}(s)\equiv \sigma_{SDD}(s)\\
		\sigma_{CDP}^{(n)}(s)&\propto \xi^{2-6(1+n)(\mu_0-1)}, \qquad \sigma_{CDP}^{(0)}(s)\equiv \sigma_{CDP}(s)		 
	\end{aligned}
\end{equation}

\section*{Conclusion}

We have considered an alternative approach to solving the problem of s-channel unitarity bounds (in particular the Finkelstein-Kajantie paradox) on diffraction production amplitudes in the Froissaron model, in which $ \sigma_{tot} (s) \propto \ln^2 (s / s_0) $. Our main assumption concerns the vertex of the three froissarons interaction, which in our approach depends on the angular and spatial momenta of the froissarons. The basic requirements for the properties of a 3f vertex are formulated and a model is constructed in which the corrections to the Dyson-Schwinger reggeon equations for $ s \to \infty $ are small in the region of small angular and spatial momenta of froissarons at the vertex. In this paper, the Dyson-Schwinger equations are considered in the leading approximation, in which only 3f-vertices are taken into account. We believe that the corrections with higher order vertices will be even smaller in this case. They will be considered in a separate work.

The constructed model of a 3f-vertex contains two parameters, one of which determines the behavior of the diffraction cross sections, which grow with energy more slowly than $ \ln ^ 2 (s / s_0 $) and the smallness of corrections to the propagator and the vertex in the Dyson-Schwinger equations. The second parameter takes into account the magnitude of the corrections, which also arise from the sub-asymptotic terms in the Froissaron propagator.

The model is applied to asymptotic estimates of the contribution to the total cross section of the three main processes of diffraction hadron production (SDD, CDP, DDD) and their generalizations to an arbitrary number of produced hadron beams with large gap rapidities between them.  All these cross sections in our approach do not functionally exceed the Froissart-Martin boundary.

Thus, in the developed approach, the Finkelstein-Kajantie contradiction does not arise at asymptotic energies.

In the diffraction interaction of protons with protons and protons with antiprotons, it is necessary to take into account the contribution of odderon along with froissaron. For elastic scattering, this is done in the Froissaron and Maximal Odderon model \cite {MN-0, MN-1, MN-2}. The results of comparing the model with experimental data showed that odderon effects are visible at high energies, but remain small. The properties of differential and total $ pp $ and $ \bar pp $ cross  sections are determined by the dominant contribution of the Froissaron. We are sure that this property also holds for the processes of diffraction production in $ pp $ and $ \bar pp $ collisions at $s\to \infty $. Therefore, we think that the inclusion of the Maximal Odderon into the considered model will not produce any problem.

\section*{Acknowledgments}

We would like to thank B. Nicolescu and V. Gusynin for important and helpful discussions and comments. Research is funded by National Academy of Sciences of Ukraine (Project No. 0120U100935).

\numberwithin{equation}{section}
\appendix

\section{Modified propagator for reggeon}\label{sect:Modified propagator}

Let us consider the possibility to have a smooth behavior of $\text{Im}H^{(k)}(s,b)$ in $b$ and  to guarantee a divergence of integrals over momenta $\vec{q_i}$. For that we write a reggeon propagators in the following form
\begin{equation}\label{eq:Suppressed-propagator-1}
	G^{(k)}(\omega, q)=\dfrac{g_ke^{-Bq^2}}{(\omega^2+a^2q^2)^{3/2-k/2}}, \quad k=0,1,2
\end{equation}

\begin{equation}
	\begin{array}{{rl}}
		\text{Im}H^{(k)}(s,b)&=\dfrac{v^2g_k}{4s}\text{Im} \left (i\int\limits_0^\infty dqq J_0(bq)e^{-Bq^2}\int \dfrac{d\omega}{2\pi i}\dfrac{e^{(1+\omega) \xi'}}{(\omega^2+a^2q^2)^{3/2-k/2}}\right )\\
		&=C\text{Im} \left (i\xi'^{1-k/2}\int\limits_0^\infty dx x^{k/2}J_0(bx/a)e^{-Bx^2/a^2}J_{1-k/2}(\xi' x)\right ),\\
		C&= \dfrac{v^2g_k\sqrt{\pi}2^{-k/2}}{8 s_0a^2\Gamma(3/2-k/2)}, \quad \xi'=\xi-i\pi/2.
	\end{array}
\end{equation}
Then considering $\xi\to \infty$ we neglect  the imaginary part in $\xi'$ and make use the integral (\cite{GR}) 
\begin{equation}\nonumber
	\begin{array}{ll}
		&\int\limits_0^\infty dx x^{\lambda-1}e^{-\alpha x^2}J_\mu(\beta x)J_\nu(\gamma x)=
		\dfrac{\beta^\mu \gamma^\nu}{2^{\nu+\mu+1}\Gamma(\nu+1)}\alpha^{-(\lambda+\nu+\mu)/2}\\ &\times\sum\limits_{m=0}^\infty\dfrac{\Gamma(m+(\lambda+\mu+\nu)/2)}{m!\Gamma(m+\mu+1)}\left(-\dfrac{\beta^2}{4\alpha^2}\right)^m\\
		&\times F\left( -m,-m-\mu;\nu+1;\dfrac{\gamma^2}{\beta^2}\right ).
	\end{array}
\end{equation}
For $a\xi \ll b$ we choose ($\mu=0, \nu=1-k, \lambda=k+1, \gamma=\xi, \beta=b/a$)
\begin{equation}\nonumber
	\begin{array}{rl}
		\text{Im}H^{(k)}(s,b)&=C_1\sum\limits_{m=0}^{\infty}\dfrac{(-1)^m}{m!}F\left(-m,-m;2-k;\dfrac{\xi^2}{(b/a)^2}\right)\left(\dfrac{b^2}{4(B/a)^2} \right)^{m}\\
		&\approx C_1\sum\limits_{m=0}^{\infty}\dfrac{(-1)^m}{m!}\left(\dfrac{b^2}{4(B/a)^2} \right)^{m}=C_1\exp\left (-\dfrac{b^2}{4(B/a)^2}\right ),\\
		C_1&=C\dfrac{a^2\xi^{1-k}}{2^{2-k}\Gamma(2-k)B}.
	\end{array}
\end{equation}
The estimation is valid for any considered $k$.

For $a\xi \gg b$ we choose ($\nu=0, \mu=1-k, \lambda=k+, \alpha=B/(a)^2 , \gamma=b/a, \beta = \xi$). Then
\begin{equation}\nonumber
	\begin{array}{rl}
		\text{Im}H^{(k)}(s,b)&=C_1\sum\limits_{m=0}^{\infty}\dfrac{(-)^m}
		 {\Gamma(m+2-k)}\left(\dfrac{\xi^2}{4B^2/a^4} \right)^{m}\\
		&\times F\left(-m,-m+k-1;1;\dfrac{(b/a)^2}{\xi^2}\right)\\
		&\approx C_1\sum\limits_{m=0}^{\infty}\dfrac{(-1)^m}{\Gamma(m+2-k)}\left(\dfrac{\xi^2}{4B^2/a^4} \right)^{m}.
	\end{array}
\end{equation}

\begin{figure}[!h]
	\centering
	\includegraphics[width=0.6\linewidth]{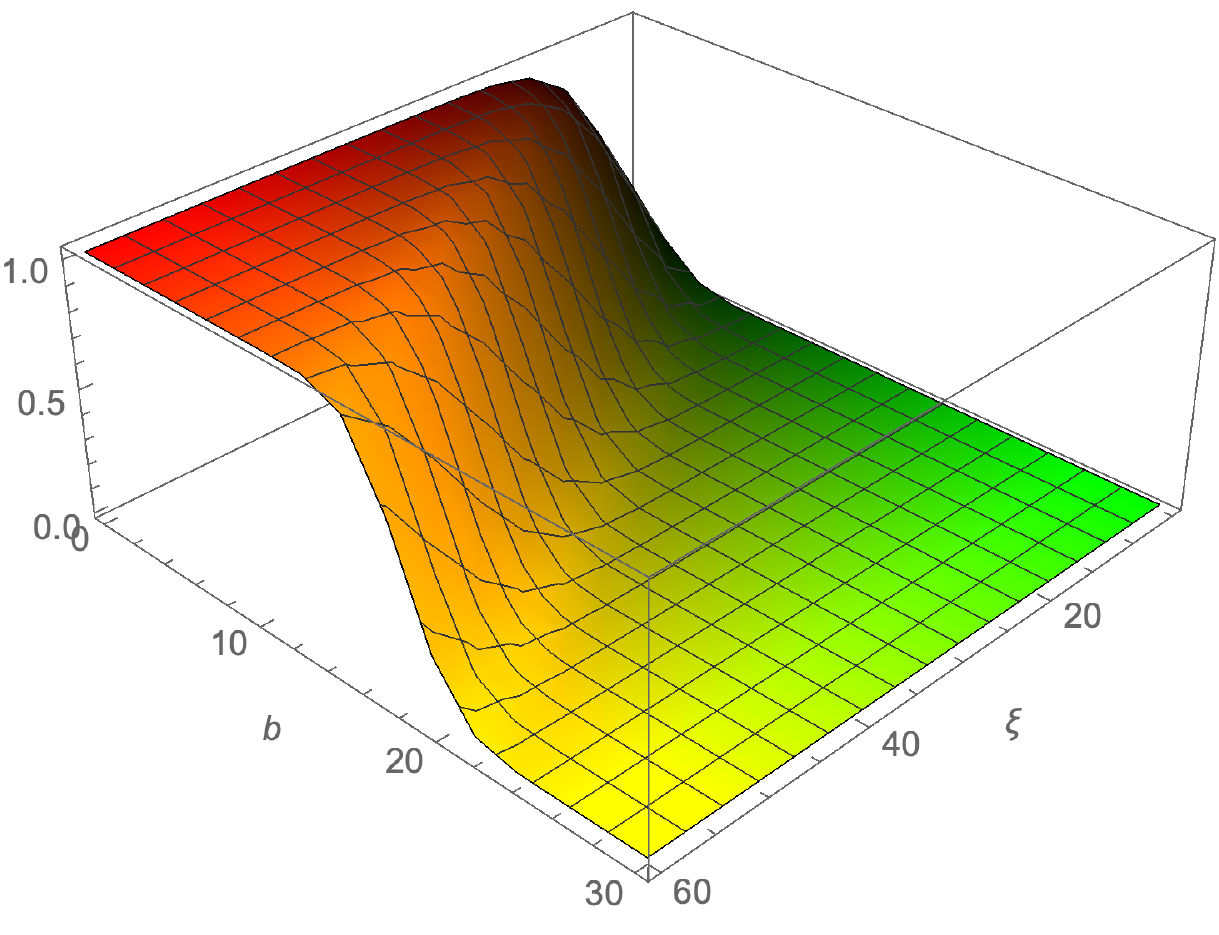}
	\caption{Amplitude $\text{Im} H^{(0)}(s,b)$ calculated for propagator \eqref{eq:Suppressed-propagator-1} at $C=1, a=0.3 \text{\,\,Gev}^{-1}, B=2 \text{\,\,Gev}^{-2}$.}
	\label{fig:H01sb}
\end{figure}

The series in the last equation can be summed separately for $k=0,1,2$ 
\begin{equation}\nonumber
	\text{Im }H^{(k)}(s,b)=\left \{
	\begin{array}{lll}
		&\dfrac{v^2g_kB}{4 s_0a^4}\left(1-e^{-a^4\xi^2/4B^2}\right), \quad &k=0,\\
		&\dfrac{v^2g_kB}{2s_0a^4}e^{-a^4\xi^2/4B^2}\text{erfi}(\xi a^2/2B), \quad &k=1,\\ 	\end{array}
	\right . 
\end{equation}
Such a suppression factor doesn't allow to calculate analytically the integrals for $H(s,b)$. However, integral for $H^{(0)}$ was calculated numerically. The results are given at Fig.~\ref{fig:H01sb} for $e^{-Bq^2}$.  At high energies, the tripole contribution $G^{(0)}(\omega,q)$ is dominating; dipole and simple pole give just the small corrections. 

Similar estimations and results one can obtain for the propagator
\begin{equation}\label{eq:Suppressed-propagator-2}
	G^{(k)}(\omega, q)=\dfrac{g_ke^{-Bq}}{(\omega^2+a^2q^2)^{3/2-k}}, \quad k=0,1,2.
\end{equation}

\section{Restrictions on the parameters $\mu_0, \lambda$}\label{sect:limits-params}
From the properties of 3f-vertex considered in the Section \ref{sect: P&3F} we have
\begin{enumerate}
\item
\begin{equation}
	\begin{aligned}
		&\Gamma_0^{(k_0,k_1,k_2)}\propto\kappa_i^\mu(k), &\\ 
		&b) &\mu(k)=\mu_0(1+\lambda k),\\
		&c)	&\mu(k)=\mu_0\dfrac{1}{1+\lambda k},\
	\end{aligned}	
\end{equation}
\item
\begin{equation}
	\mu(k)>0 \quad k=0,1,2 \quad \text{because vertex has no infinity at any k}  
\end{equation}
\item
\begin{equation}
	\begin{aligned}
	&\mu(k)<3-k  \quad k=0,1,2 \quad \text{because vertex does not cancel zero in}\\ \quad &(G_0^{(k)})^{-1}\propto \kappa^{(3-k)} .
	\end{aligned}
\end{equation}
\end{enumerate} 
Besides, we know from the smallness of corrections to main contribution (all $k=0$) that
$\mu_0>1$.
Thus
\begin{equation}
	0<\mu(k)<3-k, \quad 1<\mu_0<3.
\end{equation}
It  follows from the above inequalities that at any $k\neq 0$
\begin{equation}
	1+\lambda k >0 \Rightarrow \lambda >-1/k, \Rightarrow \lambda>-1/2.
\end{equation}
Variant b)
\begin{equation}
	0<\mu_0(1+\lambda k)<3-k, \quad 1<\mu_0<3, \Rightarrow 1<\mu_0<\dfrac{3-k}{1+\lambda k}.
\end{equation}
It follows from the last inequality that
\begin{equation}
	\dfrac{3-k}{1+\lambda k}>1, \Rightarrow \lambda <\dfrac{2-k}{k}=2/k-1
\end{equation}
The inequality for $\lambda$ should be valid for any $k$, hence $\lambda <0$.
Thus for the variant b) we have
\begin{equation}
	-1/2<\lambda <0.
\end{equation}
However, it follows from the convergence of the integrals over $\omega$ for cut reggeons at $q=0$ in evaluation of  CDP cross sections, that 
\begin{equation}
	\begin{aligned}
&		3-2\mu_0>0,  \Rightarrow \mu_0<3/2,\\
&		3-2\mu(k)-k>0, k=1, \Rightarrow 2-2\mu(1)>0, \Rightarrow \mu(1)<1, \\
&		3-2\mu(k)-k>0, k=2, \Rightarrow 1-2\mu(2)>0, \Rightarrow \mu(2)<1/2.
	\end{aligned}
\end{equation}
Hence, at $\lambda\neq 0$
\begin{equation}
	\begin{aligned}
		\mu(1)&=\mu_0(1+1\lambda)<1 \Rightarrow  1<\mu_0 <\dfrac{1}{(1+\lambda)} \Rightarrow \lambda< 0,\\	
		\mu(2)&=\mu_0(1+2\lambda)<1/2 \Rightarrow  1<\mu_0 <\dfrac{1}{2(1+2\lambda)} \Rightarrow \lambda< -1/4.
	\end{aligned}
\end{equation}
So,  the final result for $\lambda$ in the case b) is
\begin{equation}
	-1/2<\lambda< -1/4.
\end{equation}
Similarly for the variant c) 
\begin{equation}
	\lambda >0.	
\end{equation}

\section{Estimation of SDD cross section}\label{sect:SDD-cs}

To estimate $I_{SDD}(\xi)$, we rewrite it as
\begin{equation}\nonumber
	I_{SDD}(\xi)=\int\limits_{\delta}^{1-\delta}dx x^{\alpha_1-1}(1-x)^{\alpha_2-1}
\end{equation}
where $\delta=\xi_0/\xi <<1, \neq 0$ and $\alpha_1=3-\mu(k_0)-k_0>0, \quad \alpha_2=3-\mu(k_1)-\mu(k_2)-k_1-k_2$ (sign of $\alpha_2$ depends on the values of $k_i$).

Let us split the integration domain: $(\delta, \epsilon), (\epsilon,1-\epsilon), (1-\epsilon,1-\delta)$ where $\epsilon$ is constant, $\delta\ll \epsilon\ll 1$. Then
\begin{equation}\nonumber
	\begin{aligned}
		I_{SDD}(\xi)&\approx  
		=\int\limits_{\delta}^{\epsilon}dx (x^{\alpha_1-1}+x^{\alpha_2-1})+\text{C}
		=\dfrac{c_1}{\alpha_1}(\epsilon^{\alpha_1}-\delta^{\alpha_1})-\dfrac{c_2}{\alpha_2}(\epsilon^{\alpha_2}-\delta^{\alpha_2})+C\\
		&\approx	\left \{ 
		\begin{aligned}	
			&& \text{constant} \quad \text{if} \quad \alpha_2>0,\\
			&& \dfrac{c_2}{\alpha_2}(\xi/\xi_0)^{-\alpha_2},  \quad \text{if} \quad \alpha_2<0
		\end{aligned}	
		\right .
	\end{aligned}	
\end{equation}
where C is a constant.

We have obtained in the Section \ref{sect: P&3F} that a smallness of corrections in the DS equations requires $\mu_0>1$. Therefore, the following inequalities should be satisfied for the cases b) and c) from the Eq. \eqref{eq:vertex-variants} 
\begin{equation}\label{eq:mu0lambda}
	\begin{aligned}
		&\mu_1(k)=\mu(k)+k, \quad  \text{if}\,\, \lambda \neq 0 \quad \text{then}\,\, 1+\lambda k>0\,\, \text{and}\,\, \lambda>-1/2, \\
		\rm{a)} \quad &1< \mu_0<3,\\
		\rm{b)} \quad &1< \mu_0<\dfrac{3-k}{1+\lambda k},\\
		\rm{c)} \quad &1< \mu_0<(3-k)(1+\lambda k)
	\end{aligned}	
\end{equation}
where $k=0,1,2$.
Whence, taking into account inequalities  $\mu(k)>0, \mu_0>1$ and consequently $1+\lambda k>0$, it follows that 
\begin{equation}\label{eq:lambda}
	\begin{aligned}
		&\rm{b)} \quad 1<\mu_0<\frac{1}{1+2\lambda}, \quad -1/2< \lambda<0,\\
		&\rm{c)} \quad 1<\mu_0<1+2\lambda, \quad \phantom{|-1/2 <} \lambda>0.
	\end{aligned}	
\end{equation}
Thus, with obtained restrictions on $\lambda $ we have
\begin{equation}
		\sigma_{SDD}(s) \propto 
		\xi^{2-3(\mu_0-1)}\times
		\left \{
\begin{aligned}
	b) &\quad  \xi^{\lambda \mu_0 (k_0+k_1+k_2)},\,\, &-1/2<\lambda <0,\\
	c)  &\quad \xi^{-\lambda \mu_0\sum\limits_{i=0}^{2}\dfrac{k_i}{1+\lambda k_i}}, &\lambda >0.
\end{aligned}  %
\right .
\end{equation}


\begin{thebibliography}{99}
	\bibitem{Verdiev}
	I.A. Verdiev, O.V. Kanchelli, S.G. Matinyan, A.M. Popova, K.A. Ter-Martirosyan, Sov. Phys. JETP {\bf 19} (1964), 1148.
	\bibitem{Finkelstein}
	J. Finkelstein, K. Kajantie, Multiple Pomeranchuk exchange violates unitarity, Phys. Lett. {\bf 26 B} (1968), 305; Total cross-section for n-particle production in a multi-Regge model, Nuovo Cimento {\bf  56 A}, (1968) 659.
	\bibitem{G-M}
	V.N. Gribov, A. A. Migdal, Properties of the pomeranchuk pole and the branch cuts related to it at low momentum transfer, Sov. J. Nucl. Phys. 8 (1969) 583.
    \bibitem{Gribov}
    V.N. Gribov, The Theory of Complex Angular Momentum, Cambridge University Press (2003).
	\bibitem{Kaid-T-M}
	A.B. Kaidalov and K.A. Ter-Martirosyan, Nucl. Phys. {\bf B 75} (1974) 471.
	\bibitem{Kaidalov}
	A.B. Kaidalov, Diffractive production mechanisms, Phys. Rep.  {\bf 50} (1979) 157.
	\bibitem{Froissart} 
	M. Froissart, Asymptotic behavior and subtractions in the Mandelstam representation, Phys. Rev.{\bf 123} (1961) 1053. 
	\bibitem{Martin} A. Martin,  Extension of the Axiomatic Analyticity Domain of Scattering Amplitudes by Unitarity - I; Nuovo Cimento {\bf 42 A} (1966) 930.  
	\bibitem{Mar-Luk} 
	L. Łukaszuk, L., Martin, A. Absolute upper bounds for $\pi\pi$ scattering,  Nuovo Cimento  {\bf 52 A} (1967) 122.
	\bibitem{ABSW}
	H.D.I. Abarbanel, J.P. Bronzan, R.L. Sugar, A.R. White, Phys. Rept. {\bf 21} (1975), 119.
	\bibitem{BW} 
	R.C. Brower and J. H. Weis, Pomeron decoupling theorems, Rev. Mod. Phys. {\bf 47}, 605.
	\bibitem{Cardy}
	J.L. Cardy, General features of the reggeon calculus with $\alpha>1$, 	
	Nucl. Phys. {\bf B 75} (1974), 413.
	\bibitem{GLM}
	E.Gotsman, E.M.Levin and U.Maor,  Diffractive dissociation and eikonalization in high energy pp and ¯pp collisions,  Phys.Rev. {\bf D 49} (1994) 4321.
    \bibitem{KMR-1}	
    V.V.  Khoze, A. D. Martin, and M. G. Ryskin, High energy elastic and diffractive cross sections, Eur. Phys. J. {\bf C 74} (2014) 2756.
	\bibitem{MS} 
    E.S. Martynov, B.V. Struminsky, Unitarized model of hadronic diffractive dissociation, Phys. Rev. {\bf D 53} (1996) 1018.
    \bibitem{Mart-Ters} 	
    E. Martynov, G. Tersimonov, Multigap diffraction cross sections: Problems in eikonal methods for the Pomeron unitarization, Phys. Rev. {\bf D 101} (2020) 114003.
	\bibitem{KMR-3} 
    V.A. Khoze, A.D. Martin, M.G. Ryskin, Dynamics of diffractive dissociation, Eur.Phys.J. {\bf C  81} (2021) 2, 175.
    \bibitem{BTW}
    R.C. Brower, E. De Tar, J.H. Weis, Regge theory for multiparticle amplitudes, Phys. Rep.  {\bf 14} (1974) 257.
    \bibitem{Collins} Collins, P.D.B. An introduction to Regge theory \& high energy physics. Cambridge University press. 1977. 
	\bibitem{Cheng}
    H. Cheng, T.T. Wu, Limit of Cross Sections at Infinite Energy, Phys. Rev. Lett. {\bf 24} (1970) 1456.
    \bibitem{T-M}
    K.A. Ter-Martirosyan, The "quasi-eikonal" approximation, Pisma Zh. Eksp. Teor. Fiz. {\bf 15} (1972) 734.
    \bibitem{T-T}
    S. M. Troshin, N. E. Tyurin, Unitarity at the LHC energies, Phys. Part. Nucl. {\bf 35} (2004) 555.
    \bibitem{CSP}
    J.-R. Cudell,  E. Predazzi, O. V. Selyugin, New analytic unitarization schemes, Phys. Rev. {\bf D 79} (2009), 034033.
	\bibitem{DKLT-M}
    M.S. Dubovikov, B. Z. Kopellovich,  L.I. Lapidus, K.A. Ter-Martirosyan, Nuclear Physics {\bf B 123} (1977) 147.	
	\bibitem{DT-M}K.A. Ter-Martirosyan, M.S. Dubovikov, Theory of froissaron exchange,  
	Nuclear Physics, {\bf B 124} (1977) 163-188. doi: 10.1016/0550-3213(77)90283-8
	\bibitem{Ball}
	James S. Ball, An exact solution to reggeon	calculus, Nucl. Phys., {\bf B 102} (1976) 347.
    \bibitem{MN-0} 	
    E. Martynov, B. Nicolescu, Did TOTEM experiment discover the Odderon?, Phys. Lett. {\bf B 778} (2018) 414 .
    \bibitem{MN-1} 
    E. Martynov, B. Nicolescu, Evidence for maximality of strong interactions from LHC forward data, Phys. Lett. {\bf B 786} (2018) 207 .  
    \bibitem{MN-2}  
    E. Martynov, B. Nicolescu, Odderon effects in the differential cross-sections at Tevatron and LHC energies, Eur. Phys. J. {\bf C 79} (2019) 461.
	\bibitem{GR}
	I.S. Gradshteyn and I.M. Ryzhik, Table of Integrals, Series, and Products. 2007, Elsevier Inc.
\end{thebibliography}
\end{document}